\documentclass[aps,prd,twocolumn,nofootinbib,superscriptaddress,showkeys]{revtex4}
\usepackage{amssymb,amsmath,graphicx,enumerate}
\usepackage{subfigure}
\usepackage{dcolumn,booktabs}

\usepackage[dvipdfm,colorlinks=true,citecolor=blue,pdfstartview=FitH]{hyperref}

\def\bea{\begin{equation}}
\def\eea{\end{equation}}
\newcommand{\bcts}{bottom-charm tetraquarks}
\newcommand{\bct}{bottom-charm tetraquark}
\newcommand{\bctf}{$(bq)(\bar{c}\bar{q}')$}
\newcommand{\cbtf}{$(cq)(\bar{b}\bar{q}')$}
\newcommand{\hbts}{bottom-charm tetraquarks}

\newcommand{\twh}{tetraquarks with open bottom and charm}

\newcommand{\rt}{Regge trajectory}
\newcommand{\rts}{Regge trajectories}
\newcommand{\tr}{trajectory}
\newcommand{\trs}{trajectories}
\newcommand{\bfr}{{\bf r}}
\newcommand{\bfp}{{\bf p}}
\newcommand{\bfpa}{{|\bf p|}}
\newcommand{\gev}{{\rm GeV}}

\newcommand{\cltb}{$\bar{3}_c$}
\newcommand{\cltba}{\bar{3}_c}

\begin{document}
\title{$\lambda$ and $\rho$ Regge trajectories for bottom-charm tetraquarks $(bq)(\bar{c}\bar{q}')$ and $(cq)(\bar{b}\bar{q}')$}
\author{Jiao-Kai Chen}
\email{chenjk@sxnu.edu.cn, chenjkphy@outlook.com (corresponding author)}
\affiliation{School of Physics and Electronic Engineering, Shanxi Normal University, Taiyuan 030031, China}
\author{He Song}
\email{songhe\_22@163.com}
\affiliation{School of Physics and Electronic Engineering, Shanxi Normal University, Taiyuan 030031, China}
\author{Xin-Ru Liu}
\email{1170394732@qq.com}
\affiliation{School of Physics and Electronic Engineering, Shanxi Normal University, Taiyuan 030031, China}

\begin{abstract}
Using the newly proposed tetraquark Regge trajectory relations, we investigate three series of Regge trajectories for bottom-charm tetraquarks $(bq)(\bar{c}\bar{q}')$ and $(cq)(\bar{b}\bar{q}')$ with $q,q'=u,d,s$: the $\rho_1$-, $\rho_2$-, and $\lambda$-trajectories. We provide rough estimates for the masses of the $\rho_1$-, $\rho_2$-, and $\lambda$-excited states.
Except for the $\lambda$-trajectories, the complete forms of the other two series of Regge trajectories for bottom-charm tetraquarks are lengthy and cumbersome. We show that the $\rho_1$- and $\rho_2$-trajectories cannot be obtained by simply imitating meson Regge trajectories, because mesons have no substructures. To derive these trajectories, the tetraquarks' structure and substructure must be taken into consideration. Otherwise, the $\rho_1$- and $\rho_2$-trajectories would have to rely solely on fitting existing theoretical results or future experimental data. Consequently, the fundamental relationship between the slopes of the obtained trajectories and string tension would become unobvious, and the predictive power of the Regge trajectories would be compromised.
Moreover, we show that the lengthy complete forms of the $\rho_1$- and $\rho_2$-trajectories can be well approximated by simple fitted formulas. For the bottom-charm tetraquarks $(bq)(\bar{c}\bar{q}')$ and $(cq)(\bar{b}\bar{q}')$, $\rho_1$- and $\rho_2$-trajectories exhibit a behavior of $M{\sim}x^{1/2}$  $(x=n_{r_1},n_{r_2},l_1,l_2)$, whereas $\lambda$-trajectories exhibit a behavior of $M{\sim}x^{2/3}$ $(x=N_{r},L)$.
All three series of trajectories display concave downward behavior in the $(M^2,\,x)$ plane when the confining potential is linear. This conclusion holds irrespective of whether light-quark masses are included, owing to the large masses of the heavy quarks.
\end{abstract}

\keywords{$\lambda$-trajectory, $\rho$-trajectory, tetraquark, mass}
\maketitle


\section{Introduction}
The {\bcts} {\bctf} and {\cbtf} $(q,q'=u,\,d,\,s)$ have attracted significant interest, as discussed in Refs.  \cite{Jaffe:2004ph,Lipkin:2007cg,Silvestre-Brac:1993wyf,
Zouzou:1986qh,Karliner:2013dqa,
Chen:2013aba,Albuquerque:2012rq,Ortega:2020uvc,Ebert:2007rn,Zhang:2009vs,
Agaev:2016dsg,Wang:2019xzt,Wu:2018xdi,Wang:2020jgb,Ozdem:2024rrg}, although they have not yet been observed experimentally.
The masses of these {\bcts} {\bctf} and {\cbtf} have been calculated using the constituent quark model \cite{Ortega:2020uvc}, the chromomagnetic interaction model \cite{Wu:2018xdi}, the relativistic quark model \cite{Ebert:2007rn}, QCD sum rules \cite{Zhang:2009vs,Agaev:2016dsg,Wang:2019xzt,Wang:2020jgb,Ozdem:2024rrg}, and related approaches.

{\rts}\footnote{A {\rt} of bound states is generally expressed as $M=m_R+\beta_x(x+c_0)^{\nu}$ $(x=l,\,n_r)$ \cite{Chen:2022flh,Chen:2021kfw}, where $M$ is the mass of the bound state, $l$ is the orbital angular momentum, and $n_r$ is the radial quantum number. $m_R$ and $\beta_x$ are parameters. For simplicity, plots in the $(M,\,x)$ plane \cite{Xie:2024lfo}, $(M-m_R,\,x)$ plane \cite{Chen:2023cws}, $(M,\,(x+c_0)^{\nu})$ plane \cite{Burns:2010qq,Song:2025cla}, $(M^2,\,x)$ plane \cite{Chen:2018nnr}, $((M-m_R)^2,\,x)$ plane \cite{Chen:2023djq,Chen:2023web} or $((M-m_R)^{1/{\nu}},\,x)$ plane \cite{Xie:2024dfe}, are all commonly referred as Chew-Frautschi plots. {\rts} can be plotted in these various planes. }
are among the effective approaches widely used in studies of hadron spectra \cite{Burns:2010qq,Regge:1959mz,Chew:1962eu,Nambu:1978bd,Gross:2022hyw,Brodsky:2006uq,Nielsen:2018uyn,
brau:04bs,Brisudova:1999ut,Guo:2008he,Ebert:2009ub,Irving:1977ea,Collins:1971ff,
Inopin:1999nf,Afonin:2014nya,MartinContreras:2020cyg,Sergeenko:1994ck,Veseli:1996gy,
Ghasempour:2025tdu,Wilczek:2004im,Selem:2006nd,Sonnenschein:2018fph,MartinContreras:2023oqs,
Roper:2024ovj,NaikNS:2025smw,lodha:2025snn,Vishwakarma:2024ccq,
Berman:2024owc,Lodha:2024bwn,Toniato:2025gts,Feng:2023tue}.
To our knowledge, no studies address both the $\rho$- and $\lambda$-trajectories for {\bcts} {\bctf} and $(cq)(\bar{b}\bar{q}')$. In Ref. \cite{Xie:2024dfe}, {\rts} for hidden bottom and charm tetraquark were proposed using the diquark {\rt} \cite{Chen:2023cws} together with the {\rt} relations for heavy-heavy systems \cite{Chen:2022flh,Chen:2021kfw}.
In the present work, we employ the newly proposed tetraquark {\rt} relations \cite{Xie:2024dfe} to investigate both the $\lambda$- and $\rho$-{\trs} for {\bcts}. The masses of the $\lambda$- and $\rho$-excited {\twh} are roughly estimated.
We further show that the $\rho_1$- and $\rho_2$--trajectories cannot be obtained by simply imitating meson Regge trajectories, because mesons lack internal substructure. To derive these trajectories, the tetraquarks' structure and substructure must be taken into consideration.

The paper is organized as follows: In Sec. \ref{sec:rgr}, the {\rt} relations for the {\hbts} are presented.  In Sec. \ref{sec:rts}, three series of masses and three series of {\rts} are given. The conclusions are presented in Sec. \ref{sec:conc}.

\section{{\rt} relations for the bottom-charm tetraquarks $(bq)(\bar{c}\bar{q}')$ and $(cq)(\bar{b}\bar{q}')$}\label{sec:rgr}
In this section, using the diquark {\rts} \cite{Chen:2023cws}, we present {\rt} relations for bottom-charm tetraquarks $(bq)(\bar{c}\bar{q}')$ and $(cq)(\bar{b}\bar{q}')$ \cite{Xie:2024dfe}, and these relations can be used to discuss both the $\lambda$- and $\rho$-trajectories.

\subsection{Preliminary}\label{subsec:prelim}

\begin{figure}[!phtb]
\centering
\includegraphics[width=0.25\textheight]{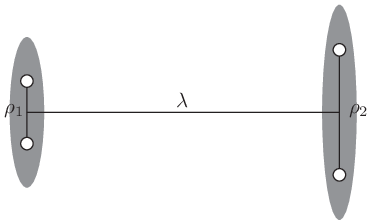}
\caption{Schematic diagram of a tetraquark in the diquark-antidiquark picture.}\label{fig:tr}
\end{figure}

In the diquark picture, tetraquarks consist of one diquark and one antidiquark, see Fig. \ref{fig:tr}. $\rho_1$ ($\rho_2$) separates the quarks (antiquarks) in the diquark (antidiquark), and $\lambda$ separates the diquark and the antidiquark. There exist three excited modes: the $\rho_1$-mode involves the radial and orbital excitation in the diquark, the $\rho_2$-mode involves the radial and orbital excitation in the antidiquark, and the $\lambda-$mode involves the radial or orbital excitation between the diquark and antidiquark. Consequently, there exist three series of {\rts}: $\rho_1$-, $\rho_2$-, and $\lambda$-{\trs}.

A diquark $(q_1q_2)$ can couple to only two irreducible color representations: $3_c\otimes3_c=\cltba\oplus{6}_c$. The $\bar{3}_c$ is the attractive channel, while in the $6_c$ representation, the internal interaction between the $q_1q_2$ pair is repulsive.
Only the {\cltb} representation of $SU_c(3)$ is considered in the present work \cite{Brodsky:2014xia,Galkin:2023wox}. The color-singlet tetraquarks under cosideration are composed of a diquark in $\bar{3}_c$ and an antidiquark in $3_c$.

In the diquark picture, the state of tetraquark is denoted as
\bea\label{tetnot}
\left((q_1q'_1)^{{\bar{3}_c}}_{n_1^{2s_1+1}l_{1j_1}}
(\bar{q}_2\bar{q}'_2)^{{{3}_c}}_{n_2^{2s_2+1}l_{2j_2}}\right)^{1_c}_{N^{2s_{3}+1}L_J},
\eea
where {\cltb} denotes the color antitriplet state of diquark, and $1_c$ represents the color singlet state of tetraquark. [The superscript $1_c$ is often omitted because the tetraquarks are colorless.] For simplicity, the notation in Eq. (\ref{tetnot}) is rewritten as $|n_1^{2s_1+1}l_{1j_1},n_2^{2s_2+1}l_{2j_2},N^{2s_3+1}L_J\rangle$. The diquark $(q_1q'_1)$ is $\{q_1q'_1\}$ or $[q_1q'_1]$. The antidiquark $(\bar{q}_2\bar{q}'_2)$ is $\{\bar{q}_2\bar{q}'_2\}$ or $[\bar{q}_2\bar{q}'_2]$. $\{qq'\}$ and $[qq']$ indicate the permutation symmetric and antisymmetric flavor wave functions, respectively. $N=N_{r}+1$, where $N_{r}=0,\,1,\,\cdots$. $n_{1,2}=n_{r_{1,2}}+1$, where $n_{r_{1,2}}=0,\,1,\,\cdots$. $N_r$, $n_{r_1}$ and $n_{r_2}$ are the radial quantum numbers of the tetraquark, diquark 1, antidiquark 2, respectively.
$\vec{J}=\vec{L}+\vec{s}_3$, $\vec{s}_3=\vec{j}_1+\vec{j}_2$, $\vec{j}_1=\vec{s}_1+\vec{l}_1$, $\vec{j}_2=\vec{s}_2+\vec{l}_2$.
$\vec{J}$, $\vec{j}_1$ and $\vec{j}_2$ are the spins of tetraquark, diquark 1, antidiquark 2, respectively. $L$, $l_{1}$ and $l_2$ are the orbital quantum numbers of tetraquark, diquark 1 and antidiquark 2, respectively. $\vec{s}_{1}$ ($\vec{s}_{2}$) is the summed spin of quarks (antiquarks) in the diquark (antidiquark). $\vec{s}_3$ is the summed spin of diquarks and antidiquarks in the tetraquark.

Evidently, the masses of the state $|n_1^{2s_1+1}l_{1j_1},n_2^{2s_2+1}l_{2j_2},N^{2s_3+1}L_J\rangle$ of the bottom-charm tetraquarks $(bq)(\bar{c}\bar{q}')$ are equal to those of the state $|n_2^{2s_2+1}l_{2j_2},n_1^{2s_1+1}l_{1j_1},N^{2s_3+1}L_J\rangle$ of $(cq')(\bar{b}\bar{q})$. Therefore, only the bottom-charm tetraquarks $(bq)(\bar{c}\bar{q}')$ are discussed. We assume $m_u=m_d$; therefore, the masses for diquarks $(bd)$ and $(cd)$ are equal to those of $(bu)$ and $(cu)$, respectively. An analogous relation holds for the tetraquark masses as well.

\subsection{Spinless Salpeter equation}
The spinless Salpeter equation \cite{Godfrey:1985xj,Ferretti:2019zyh,Bedolla:2019zwg,Durand:1981my,Durand:1983bg,Lichtenberg:1982jp,Jacobs:1986gv} reads as
\begin{eqnarray}\label{qsse}
M\Psi_{d,t}({\bfr})=\left(\omega_1+\omega_2\right)\Psi_{d,t}({\bfr})+V_{d,t}\Psi_{d,t}({\bfr}),
\end{eqnarray}
where $M$ is the bound state mass (diquark or tetraquark). $\Psi_{d,t}({\bfr})$ are the diquark wave function and the tetraquark wave function, respectively. $V_{d,t}$ denotes the diquark potential and the tetraquark potential, respectively (see Eq. (\ref{potv})). $\omega_1$ is the relativistic energy of constituent $1$ (quark or diquark), and $\omega_2$ is of constituent $2$ (quark or antidiquark),
\bea\label{omega}
\omega_i=\sqrt{m_i^2+{\bf p}^2}=\sqrt{m_i^2-\Delta}\;\; (i=1,2).
\eea
$m_1$ and $m_2$ are the effective masses of constituent $1$ and $2$, respectively.

Following Refs. \cite{Ferretti:2019zyh,Bedolla:2019zwg,Ferretti:2011zz,Eichten:1974af}, we employ the potential
\begin{align}\label{potv}
V_{d,t}&=-\frac{3}{4}\left[V_c+{\sigma}r+C\right]
\left({\bf{F}_i}\cdot{\bf{F}_j}\right)_{d,t},
\end{align}
where $V_c\propto{1/r}$ is a color Coulomb potential or a Coulomb-like potential due to one-gluon-exchange. $\sigma$ is the string tension. $C$ is a fundamental parameter \cite{Gromes:1981cb,Lucha:1991vn}. The part in the bracket is the Cornell potential \cite{Eichten:1974af}. ${\bf{F}_i}\cdot{\bf{F}_j}$ is the color-Casimir,
\bea\label{mrcc}
\langle{(\bf{F}_i}\cdot{\bf{F}_j})_{d}\rangle=-\frac{2}{3},\quad
\langle{(\bf{F}_i}\cdot{\bf{F}_j})_{t}\rangle=-\frac{4}{3}.
\eea

The spinless Salpeter equation (\ref{qsse}), together with potential (\ref{potv}), can compute masses of mesons and diquarks \cite{Ferretti:2019zyh,Godfrey:1985xj}. In the diquark picture, a tetraquark comprises a diquark in $\bar{3}_c$ and an antidiquark in $3_c$. Therefore, a tetraquark is a meson-like system, where a meson comprises a quark in $3_c$ and an antiquark in $\bar{3}_c$ \cite{Ferretti:2019zyh,Brodsky:2014xia}.
However, the diquark and antidiquark--being the constituents of tetraquarks--have finite size because they are bound states of quarks or antiquarks, whereas the quark and antiquark--the constituents of mesons and diquarks--are pointlike.
The finite-size effect of the diquark is treated differently in the literature. In Ref. \cite{Faustov:2021hjs}, the size of diquark is taken into account through appropriate form factors, whereas in Refs. \cite{Ferretti:2019zyh,Lundhammar:2020xvw} the diquark is treated as pointlike.
It is expected that the diquark's finite-size effect does not affect the {\rt} behavior but does make solving process more complex, and sometimes even intractable. In the present work, we focus on tetraquark {\rts}. For simplicity, the diquark and antidiquark are treated as pointlike, and the potential (\ref{potv}) used for mesons and diquarks is applied directly to tetraquarks.
As in the meson case--where a simple Regge trajectory relation can produce results agreeing well with experimental data and theoretical predictions--it is surprising that the present results agree well with other theoretical approaches (see Table \ref{tab:masscomp}), even though the diquark and antidiquark are treated as pointlike and the parameters are fitted to mesons and baryons.

\subsection{{\rt} relations}
When a bound state contains a light quark, the effects of relativistic kinematics--which require a nonlocal kinetic energy operator--together with momentum-dependent corrections to the potential energy operator--which introduce nonlocal modifications of the relative coordinate and adjustments to the strengths of various parts of the potential \cite{Godfrey:1985xj,Stanley:1980zm,
Capstick:1986ter,Lichtenberg:1987ms}--become significant. The kinetic-energy operator in Eq. (\ref{omega}), in configuration space, renders it difficult to arrive at rigorous analytical solutions, whether the potential is the simple Coulomb potential or a complex one \cite{Fulcher:1994ek,Lucha:1996cp,
durand:1990mm}. The {\rt} relation usually takes a simple form--for example, the well-known linear form $M^2=\alpha{J}+\beta$ for the light mesons. The {\rt} relation is not an exact solution to Eq. (\ref{qsse}) but a good approximation that reveals the relationship between the masses of bound states and their quantum numbers--$J$ (the spin of the bound state), $l$ (the orbital quantum number), and $n_r$ (the radial quantum number). The Bohr-Sommerfeld quantization approach \cite{Brau:2000st} provides, for the Cornell potential, analytical formulae for the energy spectra which closely approximate numerical exact calculations performed with the Schr\"{o}dinger or the spinless Salpeter equations. In this subsection, we simplify Eq. (\ref{qsse}) for different systems and obtain the corresponding {\rt} relations using the Bohr-Sommerfeld quantization approach.

For the heavy-heavy systems, $m_{1},m_2{\gg}{\bfpa}$, Eq. (\ref{qsse}) reduces to
\begin{eqnarray}\label{qssenrr}
M\Psi_{d,t}({\bfr})&=&\left[(m_1+m_2)+\frac{{\bfp}^2}{2\mu}\right]\Psi_{d,t}({\bfr})\nonumber\\
&&+V_{d,t}\Psi_{d,t}({\bfr}),
\end{eqnarray}
where
\bea\label{rdmu}
\mu=m_1m_2/(m_1+m_2).
\eea
By employing the Bohr-Sommerfeld quantization approach \cite{Brau:2000st} and using Eqs. (\ref{potv}) and (\ref{qssenrr}), we obtain the parameterized relation  \cite{Chen:2022flh,Chen:2021kfw}
\bea\label{massform}
M=m_R+\beta_x(x+c_{0x})^{2/3}\,\,(x=l,\,n_r,\,L,\,N_r),
\eea
with
\bea\label{parabm}
\beta_x=c_{fx}c_xc_c,\quad m_R=m_1+m_2+C',
\eea
where
\bea\label{cprime}
C'=\left\{\begin{array}{cc}
C/2, & \text{diquarks}, \\
C, & \text{tetraquarks}.
\end{array}\right.
\eea
\bea\label{sigma}
\sigma'=\left\{\begin{array}{cc}
\sigma/2, & \text{diquarks}, \\
\sigma, & \text{tetraquarks}.
\end{array}\right.
\eea
$c_{x}$ and $c_c$ are
\bea\label{cxcons}
c_c=\left(\frac{\sigma'^2}{\mu}\right)^{1/3},\quad c_{l,L}=\frac{3}{2},\quad c_{n_r,N_r}=\frac{\left(3\pi\right)^{2/3}}{2}.
\eea
$c_{fx}$ are equal theoretically to one and are fitted in practice.
In Eq. (\ref{massform}), $m_1$, $m_2$, $c_x$ and $\sigma$ are universal for the heavy-heavy systems. $c_{0x}$ vary with different {\rts}.

For the heavy-light systems ($m_1\to\infty$ and $m_2\to0$), Eq. (\ref{qsse}) simplifies to
\begin{eqnarray}\label{qssenr}
M\Psi_{d,t}({\bfr})=\left[m_1+{\bfpa}+V_{d,t}\right]\Psi_{d,t}({\bfr}).
\end{eqnarray}
By employing the Bohr-Sommerfeld quantization approach \cite{Brau:2000st} and using Eq. (\ref{qssenr}), the parameterized formula can be written as \cite{Chen:2022flh,Chen:2021kfw}
\bea\label{rtmeson}
M=m_R+\beta_x\sqrt{x+c_{0x}}\;(x=l,\,n_r,\,L,\,N_r).
\eea
$\beta_x$ is in Eq. (\ref{parabm}), and
with
\bea\label{cxcons}
c_{c}=\sqrt{\sigma'},\quad c_{l,L}=2,\quad c_{n_r,N_r}=\sqrt{2\pi}.
\eea
For the heavy-light systems, the common choice of $m_R$ is \cite{Selem:2006nd,Chen:2021kfw,Veseli:1996gy}.
\bea\label{mrm1}
m_R=m_1.
\eea
When considering the mass of the light constituent, we employ the modified formula proposed in Ref. \cite{Chen:2023cws}, i.e., Eq. (\ref{rtmeson}) with $m_R$ in (\ref{parabm}), where $m_2$ is the light constituent's mass. Eq. (\ref{rtmeson}) with (\ref{parabm}) is an extension of $M=m_1+m_2+\sqrt{a(n_r+{\alpha}l+b)}$ \cite{Afonin:2014nya} and $(M-m_1-m_2-C)^2=\alpha_x(x+c_0)^{\gamma}$ \cite{Chen:2022flh}.

\begin{table}[!phtb]
\caption{Coefficients for heavy-heavy systems (HHS) and heavy-light systems (HLS).}  \label{tab:eparam}
\centering
\begin{tabular*}{0.47\textwidth}{@{\extracolsep{\fill}}ccc@{}}
\hline\hline
                   & HHS &  HLS   \\
\hline
$\nu$    & $2/3$ & $1/2$   \\
$c_c$    & $\left({\sigma'^2}/{\mu}\right)^{1/3}$    & $\sqrt{\sigma'}$  \\
$c_{l,\,L}$    & $3/2$ & $2$   \\
$c_{n_r,\,N_r}$ & ${\left(3\pi\right)^{2/3}}/{2}$      & $\sqrt{2\pi}$   \\
\hline
\hline
\end{tabular*}
\end{table}

When Eq. (\ref{rtmeson}) with (\ref{parabm}) is employed to discuss the heavy-light systems, and Eq. (\ref{massform}) with (\ref{parabm}) is applied to the heavy-heavy systems, we have a general form of the {\rts} \cite{Chen:2022flh,Xie:2024lfo}
\begin{align}\label{massfinal}
M=&m_R+\beta_x(x+c_{0x})^{\nu}\,\,(x=l,\,n_r,\,L,\,N_r),\nonumber\\
m_R=&m_1+m_2+C',\quad \beta_x=c_{fx}c_xc_{c},
\end{align}
where ${\nu}$, $c_x$ and $c_{c}$ are listed in Table \ref{tab:eparam}. $c_{fx}$ are theoretically equal to one and are fitted in practice. $c_{0x}$ vary with different {\rts}. Eq. (\ref{massfinal}) can be employed to discuss various systems including the heavy-heavy systems, and the heavy-light systems: diquarks, mesons, triquarks, baryons, tetraquarks, and pentaquarks \cite{Chen:2023djq,Chen:2023web,Song:2024bkj,Song:2025cla}.

It should be noted that the general form (\ref{massfinal}) is provisional, as there are multiple methods for including the masses of the light constituents \cite{Chen:2023cws}. Consequently, additional theoretical and experimental data are required to identify the most suitable method.

\subsection{{\rt} relations for the bottom-charm tetraquarks}
The bottom-charm tetraquarks $(bq)(\bar{c}\bar{q}')$ consists of a heavy-light diquark $(bq)$ and a heavy-light antidiquark $(\bar{c}\bar{q}')$. The diquark $(bq)$ and antidiquark $(\bar{c}\bar{q}')$ are heavy because the bottom/charm quark are both heavy. Therefore, the bottom-charm tetraquarks $(bq)(\bar{c}\bar{q}')$ behave as heavy-heavy systems for the $\lambda$-mode excitations. Meanwhile, the {\rts} for the bottom-charm tetraquarks exhibit the properties of the heavy-light systems for the $\rho_1$- and $\rho_2$-mode excitations. According to Eq. (\ref{massfinal}), we list the relations for {\twh} \cite{Xie:2024dfe}
\begin{align}\label{t2qx}
M&=m_{R{\lambda}}+\beta_{x_{\lambda}}(x_{\lambda}+c_{0x_{\lambda}})^{2/3}\;(x_{\lambda}=L,\,N_r),\nonumber\\
M_{\rho_1}&=m_{R\rho_1}+\beta_{x_{\rho_1}}\sqrt{x_{\rho_1}+c_{0x_{\rho_1}}}\;(x_{\rho_1}=l_{1},\,n_{r_1}),\nonumber\\
M_{\rho_2}&=m_{R\rho_2}+\beta_{x_{\rho_2}}\sqrt{x_{\rho_2}+c_{0x_{\rho_2}}}\;(x_{\rho_2}=l_{2},\,n_{r_2}),
\end{align}
where
\begin{align}\label{pa2qQx}
m_{R{\lambda}}&=M_{\rho_1}+M_{\rho_2}+C,\nonumber\\
m_{R\rho_1}&=m_{b}+m_{q}+C/2,\nonumber\\
m_{R\rho_2}&=m_{c}+m_{q^{\prime}}+C/2,\nonumber\\
\beta_{L}&=\frac{3}{2}\left(\frac{\sigma^2}{\mu_{\lambda}}\right)^{1/3}c_{fL},\nonumber\\ \beta_{N_r}&=\frac{(3\pi)^{2/3}}{2}\left(\frac{\sigma^2}{\mu_{\lambda}}\right)^{1/3}c_{fN_r},\nonumber\\
\mu_{\lambda}&=\frac{M_{\rho_1}M_{\rho_2}}{M_{\rho_1}+M_{\rho_2}},\nonumber\\
\beta_{l_1}&=\sqrt{2\sigma}c_{fl_1},\quad \beta_{n_{r_1}}=\sqrt{\pi\sigma}c_{fn_{r_1}},\nonumber\\
\beta_{l_2}&=\sqrt{2\sigma}c_{fl_2},\quad \beta_{n_{r_2}}=\sqrt{\pi\sigma}c_{fn_{r_2}}.
\end{align}
$M$, $M_{\rho_1}$, $M_{\rho_2}$, $m_{b}$, $m_{c}$, and $m_{q}$ are the tetraquark mass, the diquark mass, the antidiquark mass, the bottom quark mass, the charm quark mass, and the light quark mass, respectively. $\sigma$ is the string tension. $C$ is a fundamental parameter. $c_{fx}$ are theoretically equal to one but are fitted in practice. $c_{0x}$ vary with {\rts}.

According to Eqs. (\ref{t2qx}) and (\ref{pa2qQx}), we have
\bea\label{summ}
M=M_{\rho_1}+M_{\rho_2}+C+\beta_{x_{\lambda}}(x_{\lambda}+c_{0x_{\lambda}})^{2/3}
\eea
when the diquark and antidiquark are regarded as constituents and their internal structures are neglected. The corresponding binding
energy is $\epsilon=C+\beta_{x_{\lambda}}(x_{\lambda}+c_{0x_{\lambda}})^{2/3}$. When the substructures of the diquark and antidiquark are considered, we have from Eqs. (\ref{t2qx}) and (\ref{pa2qQx})
\begin{align}\label{summf}
M=&m_{b}+m_{q}+m_{c}+m_{q^{\prime}}+2C+\beta_{x_{\lambda}}(x_{\lambda}+c_{0x_{\lambda}})^{2/3}\nonumber\\
&+\beta_{x_{\rho_1}}\sqrt{x_{\rho_1}+c_{0x_{\rho_1}}}+\beta_{x_{\rho_2}}\sqrt{x_{\rho_2}+c_{0x_{\rho_2}}}.
\end{align}
The binding energy in this case is $\epsilon=2C+\beta_{x_{\lambda}}(x_{\lambda}+c_{0x_{\lambda}})^{2/3}
+\beta_{x_{\rho_1}}\sqrt{x_{\rho_1}+c_{0x_{\rho_1}}}
+\beta_{x_{\rho_2}}\sqrt{x_{\rho_2}+c_{0x_{\rho_2}}}$. Eq. (\ref{summf}) clearly shows three series of {\rts}: the $\lambda$-{\trs} with $x_{\rho_1}$ and $x_{\rho_2}$ fixed; the $\rho_1$-{\trs} with $x_{\lambda}$ and $x_{\rho_2}$ fixed; and the $\rho_2$-{\trs} with $x_{\lambda}$ and $x_{\rho_1}$ fixed.

For later convenience, we refer to the {\rts} obtained from Eqs. (\ref{t2qx}) and (\ref{pa2qQx}) or from Eqs. (\ref{summf}) and (\ref{pa2qQx}) as the complete forms of the {\rts}. The obtained constant and the mode under consideration are referred to the main part of the {\rts}. For example, when considering the $\rho_1$-trajectories, $\beta_{x_{\rho_2}}\sqrt{x_{\rho_2}+c_{0x_{\rho_2}}}$ is constant, while $\beta_{x_{\lambda}}(x_{\lambda}+c_{0x_{\lambda}})^{2/3}$ becomes a function of $x_{\rho_1}$ (through the dependence in $\beta_{x_{\lambda}}$). Therefore, the main part of the $\rho_1$-trajectories is
\begin{align}\label{rtmaina}
\widetilde{m}_R+\beta_{x_{\rho_1}}\sqrt{x_{\rho_1}+c_{0x_{\rho_1}}},
\end{align}
where
\bea\label{rtmainr}
\widetilde{m}_R=m_{b}+m_{q}+m_{c}+m_{q^{\prime}}+2C
+\beta_{x_{\rho_2}}\sqrt{x_{\rho_2}+c_{0x_{\rho_2}}}.
\eea
The difference between the complete form of the $\rho_1$-{\tr} and its main part is $\beta_{x_{\lambda}}(x_{\lambda}+c_{0x_{\lambda}})^{2/3}$.

\section{{\rts} for the bottom-charm tetraquarks}\label{sec:rts}
In this section, {\rts} for bottom-charm tetraquarks are discussed. The masses of bottom-charm tetraquarks are crudely estimated.

\subsection{Parameters}
The quark masses, the string tension $\sigma$ and the parameter $C$ are from Ref. \cite{Faustov:2021hjs}. The parameters for the heavy-light diquarks $(bu)$, $(bs)$, $(cu)$, and $(cs)$ are from Ref. \cite{Chen:2023cws} and listed in Table \ref{tab:parmv}. With these parameters determined, the $\rho$-modes and the diquark masses can be discussed, see Eqs. (\ref{t2qx}) and (\ref{pa2qQx}).
To discuss the masses of the {\hbts} and the $\lambda$-modes excited states,
the parameters $c_{fL}$, $c_{fN_r}$, $c_{0{L}}$ and $c_{0N_r}$ are determined by the following relations \cite{Xie:2024dfe}
\begin{eqnarray}
c_{fL}=&1.116 + 0.013\mu_{\lambda},\; c_{0L}=0.540- 0.141\mu_{\lambda}, \label{fitcfxl}\\
c_{fN_r}=&1.008 + 0.008\mu_{\lambda}, \;  c_{0N_r}=0.334 - 0.087\mu_{\lambda},\label{fitcfxnr}
\end{eqnarray}
where $\mu_{\lambda}$ is the reduced masses, see Eq. (\ref{pa2qQx}). Eqs. (\ref{fitcfxl}) and (\ref{fitcfxnr}) are obtained when $\mu_{\lambda}<3.83$ ${\gev}$. As $\mu_{\lambda}>3.83$ ${\gev}$, the relations in Eqs. (\ref{fitcfxl}) and (\ref{fitcfxnr}) should be adjusted by fitting theoretical and experimental data.
Once all parameters are determined, the {\rts} can be discussed and can be used to estimate the masses of the {\twh}.

\begin{table}[!phtb]
\caption{Parameter values \cite{Chen:2023cws,Faustov:2021hjs}.}  \label{tab:parmv}
\centering
\begin{tabular*}{0.47\textwidth}{@{\extracolsep{\fill}}cl@{}}
\hline\hline
          & $m_{u,d}=0.33\; {\gev}$, \; $m_s=0.50\; {\gev}$, \; $m_b=4.88\; {\gev}$,  \\
          & $m_c=1.55\; {\gev}$, \; $\sigma=0.18\; {\gev^2}$,\; $C=-0.3\; {\gev}$, \\
$(bu)$    & $c_{fn_{r_1}}=0.988$,\; $c_{fl_1}=0.965$,  \\
          & $c_{0n_{r_1}}(1^1s_0)=0.125$,\;$c_{0n_{l_1}}(1^1s_0)=0.18$,\\
          & $c_{0n_{r_1}}(1^3s_1)=0.155$,\;$c_{0n_{l_1}}(1^3s_1)=0.22$,\\
$(bs)$    & $c_{fn_{r_1}}=0.953$,\; $c_{fl_1}=0.919$,  \\
          & $c_{0n_{r_1}}(1^1s_0)=0.08$,\;$c_{0n_{l_1}}(1^1s_0)=0.115$\\
          & $c_{0n_{r_1}}(1^3s_1)=0.11$,\;$c_{0n_{l_1}}(1^3s_1)=0.16$,\\
$(cu)$    & $c_{fn_{r_2}}=1.000$,\; $c_{fl_2}=1.038$,  \\
          & $c_{0n_{r_2}}(1^1s_0)=0.065$,\;$c_{0n_{l_2}}(1^1s_0)=0.095$,\\
          & $c_{0n_{r_2}}(1^3s_1)=0.17$,\;$c_{0n_{l_2}}(1^3s_1)=0.19$,\\
$(cs)$    & $c_{fn_{r_2}}=1.016$,\; $c_{fl_2}=1.015$,  \\
          & $c_{0n_{r_2}}(1^1s_0)=0.03$,\;$c_{0n_{l_2}}(1^1s_0)=0.055$,\\
          & $c_{0n_{r_2}}(1^3s_1)=0.095$,\;$c_{0n_{l_2}}(1^3s_1)=0.135$.\\
\hline
\hline
\end{tabular*}
\end{table}

\subsection{${\rho}$-{\trs}}\label{subsec:rho}

When calculating the $\rho_1$-mode excitations, the scalar antidiquark is used; the other modes are kept in their ground states, and the parameters correspond to the radial ground states of those modes. Masses of the radially and orbitally excited $\rho_1$-mode states are roughly estimated using Eqs. (\ref{t2qx}), (\ref{fitcfxl}) and (\ref{fitcfxnr}) together with the parameters in Table \ref{tab:parmv}. The calculated results are listed in Table \ref{tab:massrho}. Masses of the radially and orbitally excited $\rho_2$-mode states can be calculated in a similar manner; the results are listed in Table \ref{tab:massrhob}.

\begin{table*}[!htbp]
\caption{Masses of the $\rho_1$-mode excited states of {\bcts} (in ${\gev}$). The notation in Eq. (\ref{tetnot}) is rewritten as $|n_1^{2s_1+1}l_{1j_1},n_2^{2s_2+1}l_{2j_2},N^{2s_3+1}L_J\rangle$. Eqs. (\ref{t2qx}), (\ref{fitcfxl}) and (\ref{fitcfxnr}) are used.}  \label{tab:massrho}
\centering
\begin{tabular*}{1.0\textwidth}{@{\extracolsep{\fill}}ccccc@{}}
\hline\hline
  $|n_1^{2s_1+1}l_{1j_1},n_2^{2s_2+1}l_{2j_2},N^{2s_3+1}L_J\rangle$        & $(bu)(\bar{c}\bar{u})$  &  $(bu)(\bar{c}\bar{s})$  &  $(bs)(\bar{c}\bar{u})$ &  $(bs)(\bar{c}\bar{s})$  \\
\hline
 $|1^1s_0, 1^1s_0, 1^1S_0\rangle$  &7.17 &7.28   &7.28  &7.39  \\
 $|2^1s_0, 1^1s_0, 1^1S_0\rangle$  &7.69 &7.80   &7.82  &7.92  \\
 $|3^1s_0, 1^1s_0, 1^1S_0\rangle$  &7.99 &8.09   &8.11  &8.21  \\
 $|4^1s_0, 1^1s_0, 1^1S_0\rangle$  &8.22 &8.32   &8.33  &8.43  \\
 $|5^1s_0, 1^1s_0, 1^1S_0\rangle$  &8.41 &8.51   &8.52  &8.62  \\
\hline
 $|1^3s_1, 1^1s_0, 1^3S_1\rangle$  &7.20 &7.31   &7.32  &7.42  \\
 $|2^3s_1, 1^1s_0, 1^3S_1\rangle$  &7.71 &7.81   &7.83  &7.93  \\
 $|3^3s_1, 1^1s_0, 1^3S_1\rangle$  &8.00 &8.10   &8.12  &8.22  \\
 $|4^3s_1, 1^1s_0, 1^3S_1\rangle$  &8.22 &8.33   &8.34  &8.44  \\
 $|5^3s_1, 1^1s_0, 1^3S_1\rangle$  &8.42 &8.52   &8.52  &8.63  \\
\hline
 $|1^1s_0, 1^1s_0, 1^1S_0\rangle$  &7.16  &7.26  &7.27  &7.37  \\
 $|1^1p_1, 1^1s_0, 1^3S_1\rangle$  &7.54  &7.64  &7.66  &7.76  \\
 $|1^1d_2, 1^1s_0, 1^5S_2\rangle$  &7.76  &7.87  &7.88  &7.98  \\
 $|1^1f_3, 1^1s_0, 1^7S_3\rangle$  &7.94  &8.04  &8.05  &8.15  \\
 $|1^1g_4, 1^1s_0, 1^9S_4\rangle$  &8.09  &8.19  &8.19  &8.30  \\
 $|1^1h_5, 1^1s_0, 1^{11}S_5\rangle$&8.22 &8.32  &8.32  &8.42  \\
\hline
 $|1^3s_1, 1^1s_0, 1^3S_1\rangle$  &7.18  &7.29  &7.30  &7.40  \\
 $|1^3p_2, 1^1s_0, 1^5S_2\rangle$  &7.55  &7.65  &7.67  &7.77  \\
 $|1^3d_3, 1^1s_0, 1^7S_3\rangle$  &7.77  &7.87  &7.89  &7.99  \\
 $|1^3f_4, 1^1s_0, 1^9S_4\rangle$  &7.94  &8.05  &8.05  &8.16  \\
 $|1^3g_5, 1^1s_0, 1^{11}S_5\rangle$&8.09 &8.20  &8.20  &8.30  \\
 $|1^3h_6, 1^1s_0, 1^{13}S_6\rangle$&8.23 &8.33  &8.33  &8.43  \\
\hline\hline
\end{tabular*}
\end{table*}

\begin{table*}[!htbp]
\caption{Masses of the $\rho_2$-mode excited states of {\bcts} (in ${\gev}$). The notation in Eq. (\ref{tetnot}) is rewritten as $|n_1^{2s_1+1}l_{1j_1},n_2^{2s_2+1}l_{2j_2},N^{2s_3+1}L_J\rangle$. Eqs. (\ref{t2qx}), (\ref{fitcfxl}) and (\ref{fitcfxnr}) are used.}  \label{tab:massrhob}
\centering
\begin{tabular*}{1.0\textwidth}{@{\extracolsep{\fill}}ccccc@{}}
\hline\hline
  $|n_1^{2s_1+1}l_{1j_1},n_2^{2s_2+1}l_{2j_2},N^{2s_3+1}L_J\rangle$        & $(bu)(\bar{c}\bar{u})$  &  $(bu)(\bar{c}\bar{s})$  &  $(bs)(\bar{c}\bar{u})$ &  $(bs)(\bar{c}\bar{s})$  \\
\hline
 $|1^1s_0, 1^1s_0, 1^1S_0\rangle$  &7.17  &7.28   &7.28  &7.39  \\
 $|1^1s_0, 2^1s_0, 1^1S_0\rangle$  &7.73  &7.89   &7.84  &8.00  \\
 $|1^1s_0, 3^1s_0, 1^1S_0\rangle$  &8.02  &8.19   &8.13  &8.30  \\
 $|1^1s_0, 4^1s_0, 1^1S_0\rangle$  &8.25  &8.42   &8.35  &8.53  \\
 $|1^1s_0, 5^1s_0, 1^1S_0\rangle$  &8.44  &8.62   &8.55  &8.73  \\
\hline
 $|1^1s_0, 1^3s_1, 1^3S_1\rangle$  &7.28  &7.37   &7.39  &7.48  \\
 $|1^1s_0, 2^3s_1, 1^3S_1\rangle$  &7.76  &7.91   &7.87  &8.02  \\
 $|1^1s_0, 3^3s_1, 1^3S_1\rangle$  &8.04  &8.21   &8.15  &8.31  \\
 $|1^1s_0, 4^3s_1, 1^3S_1\rangle$  &8.27  &8.44   &8.38  &8.54  \\
 $|1^1s_0, 5^3s_1, 1^3S_1\rangle$  &8.46  &8.63   &8.57  &8.74  \\
\hline
 $|1^1s_0, 1^1s_0, 1^1S_0\rangle$  &7.17  &7.29   &7.28  &7.40  \\
 $|1^1s_0, 1^1p_1, 1^3S_1\rangle$  &7.61  &7.75   &7.72  &7.85  \\
 $|1^1s_0, 1^1d_2, 1^5S_2\rangle$  &7.85  &7.98   &7.96  &8.09  \\
 $|1^1s_0, 1^1f_3, 1^7S_3\rangle$  &8.03  &8.17   &8.14  &8.27  \\
 $|1^1s_0, 1^1g_4, 1^9S_4\rangle$  &8.19  &8.32   &8.30  &8.43  \\
 $|1^1s_0, 1^1h_5, 1^{11}S_5\rangle$&8.33 &8.46   &8.44  &8.57  \\
\hline
 $|1^1s_0, 1^3s_1, 1^3S_1\rangle$  &7.25  &7.36   &7.36  &7.47  \\
 $|1^1s_0, 1^3p_2, 1^5S_2\rangle$  &7.63  &7.77   &7.74  &7.88  \\
 $|1^1s_0, 1^3d_3, 1^7S_3\rangle$  &7.87  &7.80   &7.97  &8.11  \\
 $|1^1s_0, 1^3f_4, 1^9S_4\rangle$  &8.05  &8.18   &8.16  &8.29  \\
 $|1^1s_0, 1^3g_5, 1^{11}S_5\rangle$&8.21 &8.33   &8.31  &8.44  \\
 $|1^1s_0, 1^3h_6, 1^{13}S_6\rangle$&8.34 &8.47   &8.45  &8.58  \\
\hline\hline
\end{tabular*}
\end{table*}

\begin{figure*}[!phtb]
\centering
\subfigure[]{\label{subfigure:cfa}\includegraphics[scale=0.45]{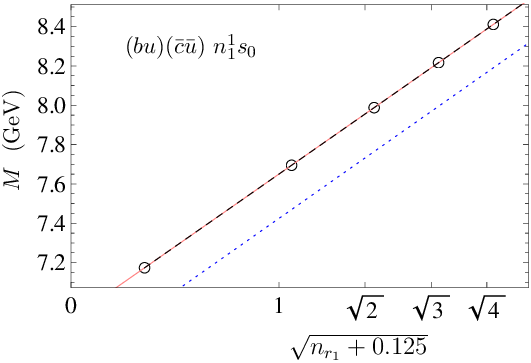}}
\subfigure[]{\label{subfigure:cfa}\includegraphics[scale=0.45]{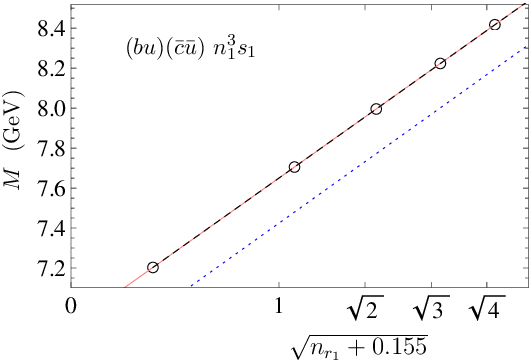}}
\subfigure[]{\label{subfigure:cfa}\includegraphics[scale=0.45]{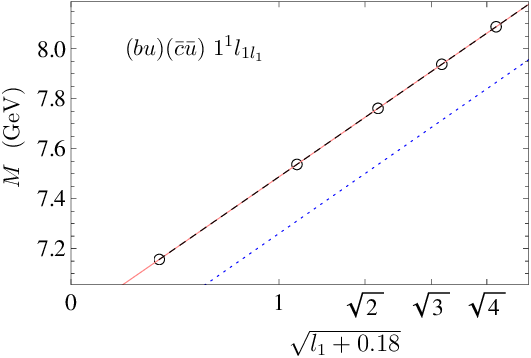}}
\subfigure[]{\label{subfigure:cfa}\includegraphics[scale=0.45]{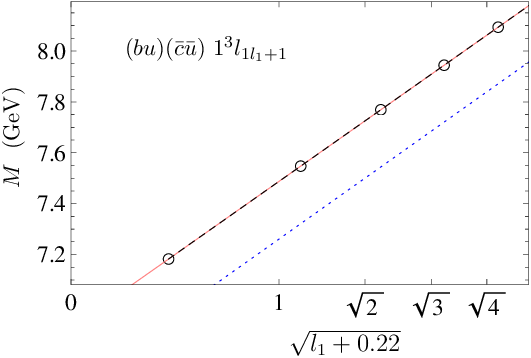}}
\subfigure[]{\label{subfigure:cfa}\includegraphics[scale=0.45]{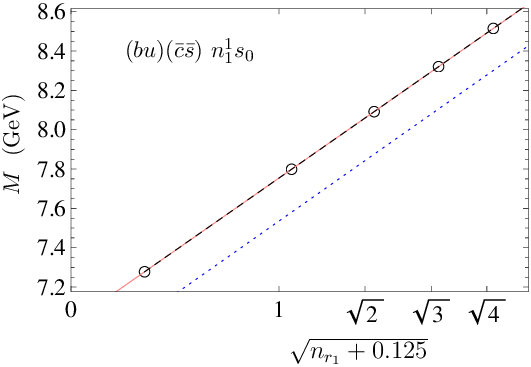}}
\subfigure[]{\label{subfigure:cfa}\includegraphics[scale=0.45]{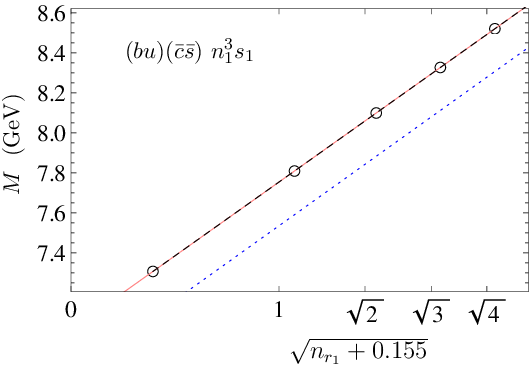}}
\subfigure[]{\label{subfigure:cfa}\includegraphics[scale=0.45]{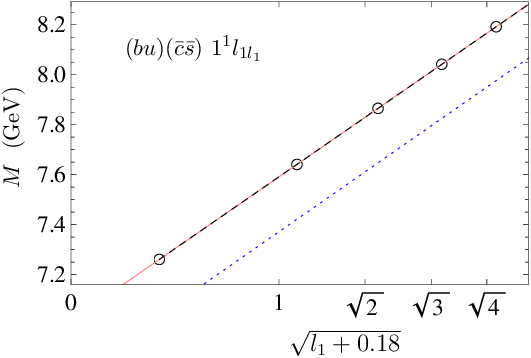}}
\subfigure[]{\label{subfigure:cfa}\includegraphics[scale=0.45]{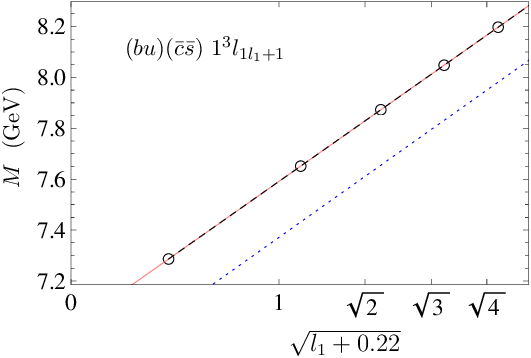}}
\subfigure[]{\label{subfigure:cfa}\includegraphics[scale=0.45]{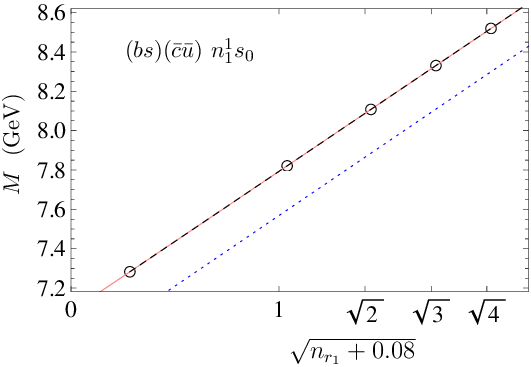}}
\subfigure[]{\label{subfigure:cfa}\includegraphics[scale=0.45]{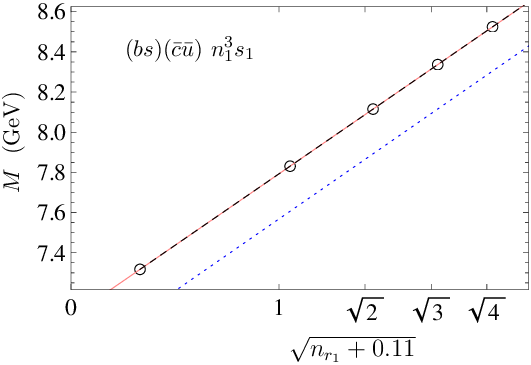}}
\subfigure[]{\label{subfigure:cfa}\includegraphics[scale=0.45]{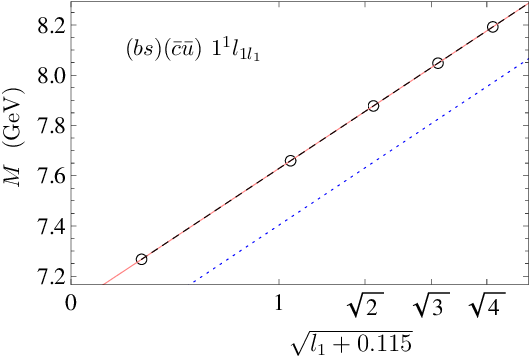}}
\subfigure[]{\label{subfigure:cfa}\includegraphics[scale=0.45]{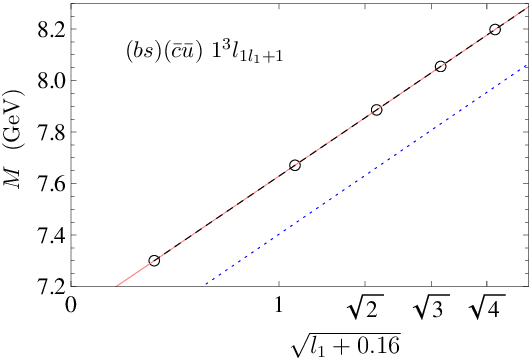}}
\subfigure[]{\label{subfigure:cfa}\includegraphics[scale=0.45]{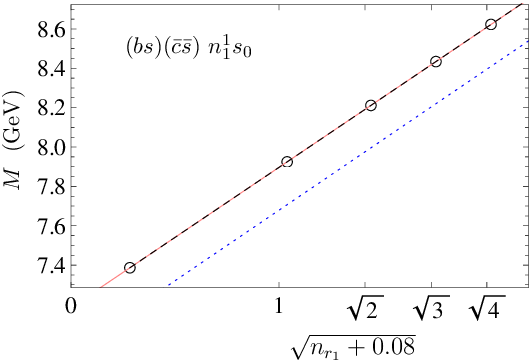}}
\subfigure[]{\label{subfigure:cfa}\includegraphics[scale=0.45]{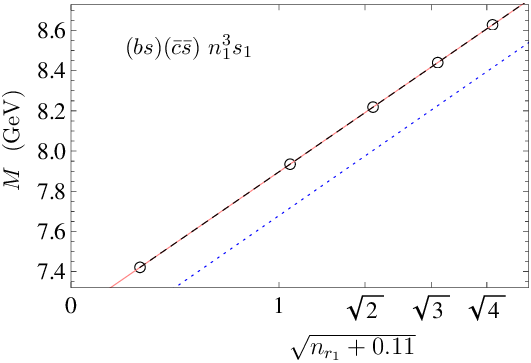}}
\subfigure[]{\label{subfigure:cfa}\includegraphics[scale=0.45]{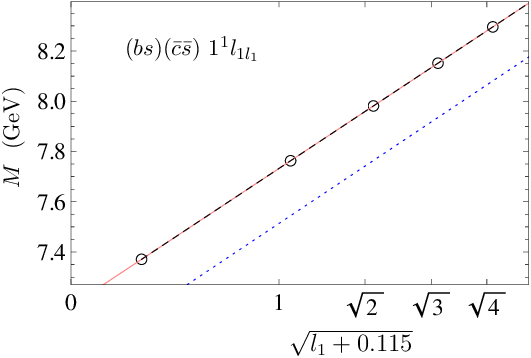}}
\subfigure[]{\label{subfigure:cfa}\includegraphics[scale=0.45]{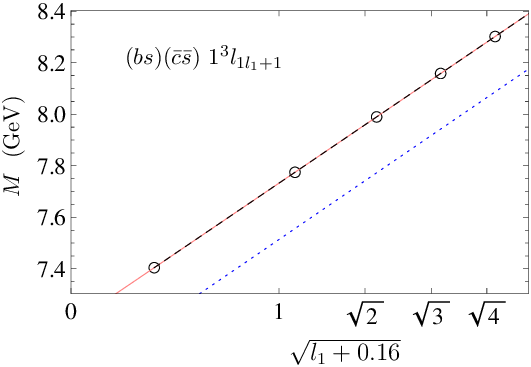}}
\caption{Orbital and radial $\rho_1$-{\trs} for tetraquarks $(bu)(\bar{c}\bar{u})$, $(bu)(\bar{c}\bar{s})$, $(bs)(\bar{c}\bar{u})$, and $(bs)(\bar{c}\bar{s})$. $n_{r_1}$ and ${l_1}$ are the radial and orbital quantum numbers for the $\rho_1$-mode, respectively. Circles represent the predicted data listed in Table \ref{tab:massrho}. The black dashed lines correspond to the complete forms of the $\rho_1$-{\trs}, obtained from Eqs. (\ref{t2qx}) and (\ref{pa2qQx}) or (\ref{summf}) and (\ref{pa2qQx}). The pink lines correspond to the fitted formulas, obtained by linearly fitting the calculated data in Table \ref{tab:massrho}; these formulas are listed in Table \ref{tab:formulas}. The blue dotted lines correspond to the main parts of the full forms, which are also listed in Table \ref{tab:formulas}. $n_1^1s_0$ and $n^3_1s_1$ denote radial {\rts} for spin-0 and spin-1 diquarks, respectively; $1^1l_{1l_1}$ and $1^3l_{1l_{1}+1}$ denote orbital {\rts} for spin-0 and spin-1 diquarks, respectively.}\label{fig:rhoa}
\end{figure*}

\begin{figure*}[!phtb]
\centering
\subfigure[]{\label{subfigure:cfa}\includegraphics[scale=0.45]{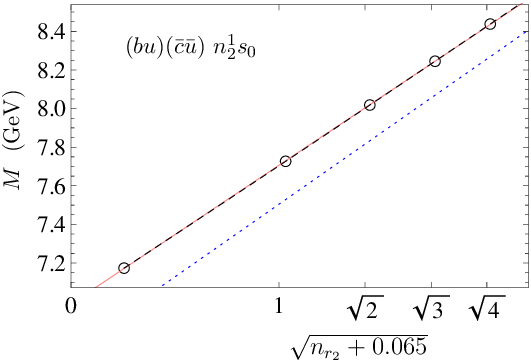}}
\subfigure[]{\label{subfigure:cfa}\includegraphics[scale=0.45]{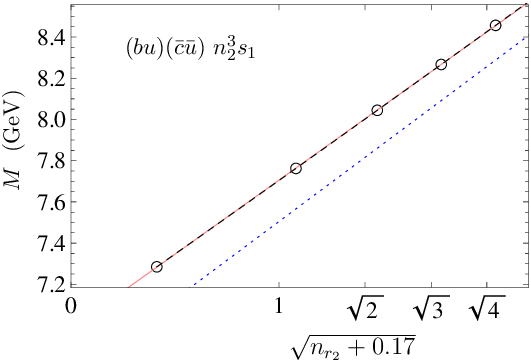}}
\subfigure[]{\label{subfigure:cfa}\includegraphics[scale=0.45]{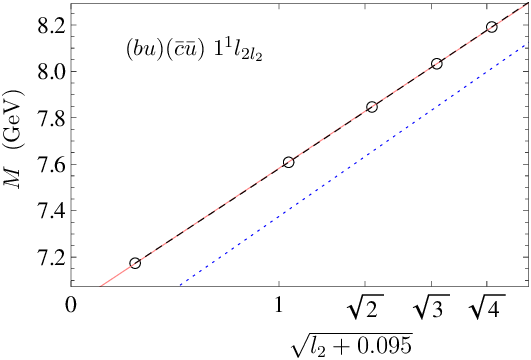}}
\subfigure[]{\label{subfigure:cfa}\includegraphics[scale=0.45]{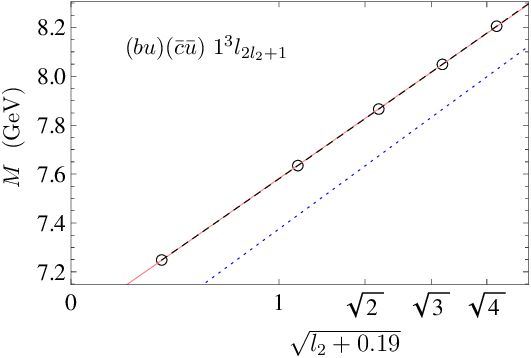}}
\subfigure[]{\label{subfigure:cfa}\includegraphics[scale=0.45]{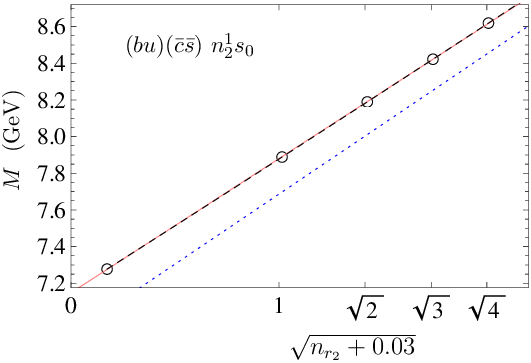}}
\subfigure[]{\label{subfigure:cfa}\includegraphics[scale=0.45]{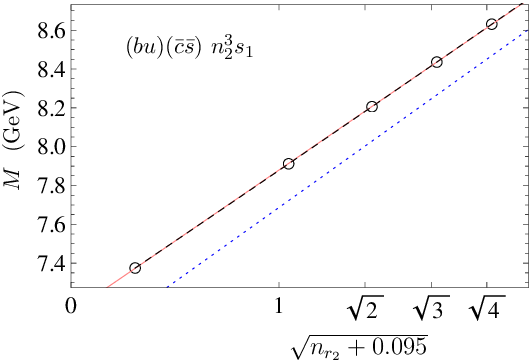}}
\subfigure[]{\label{subfigure:cfa}\includegraphics[scale=0.45]{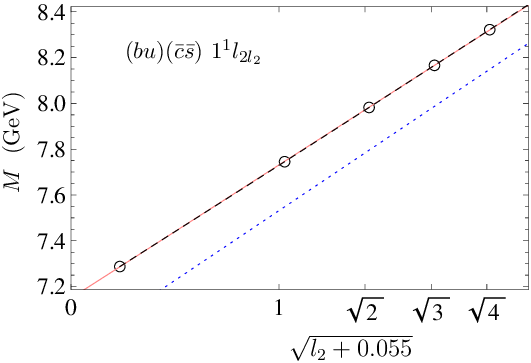}}
\subfigure[]{\label{subfigure:cfa}\includegraphics[scale=0.45]{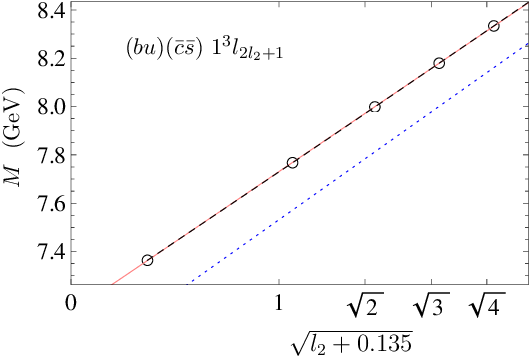}}
\subfigure[]{\label{subfigure:cfa}\includegraphics[scale=0.45]{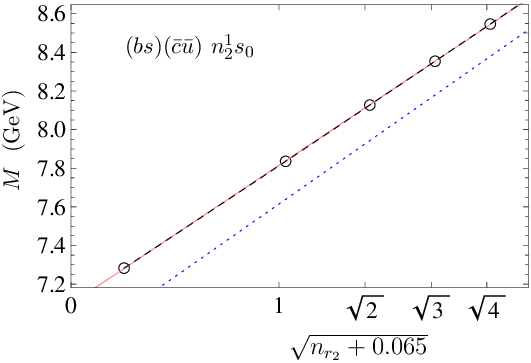}}
\subfigure[]{\label{subfigure:cfa}\includegraphics[scale=0.45]{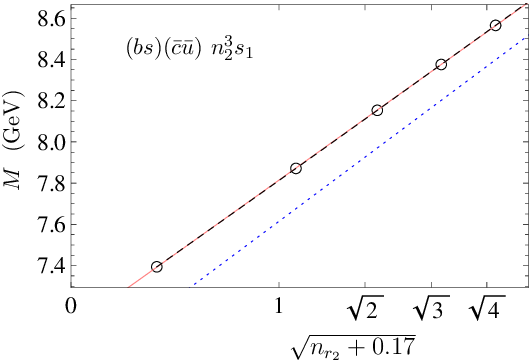}}
\subfigure[]{\label{subfigure:cfa}\includegraphics[scale=0.45]{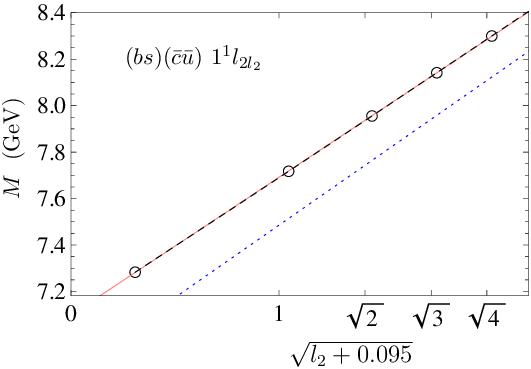}}
\subfigure[]{\label{subfigure:cfa}\includegraphics[scale=0.45]{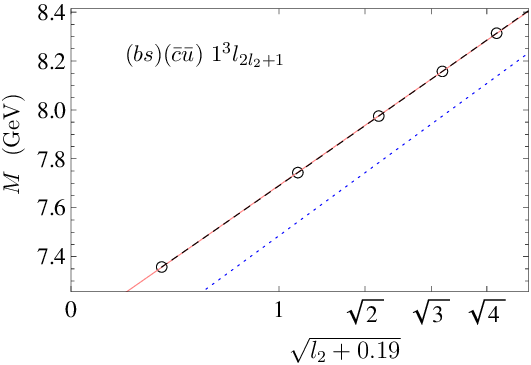}}
\subfigure[]{\label{subfigure:cfa}\includegraphics[scale=0.45]{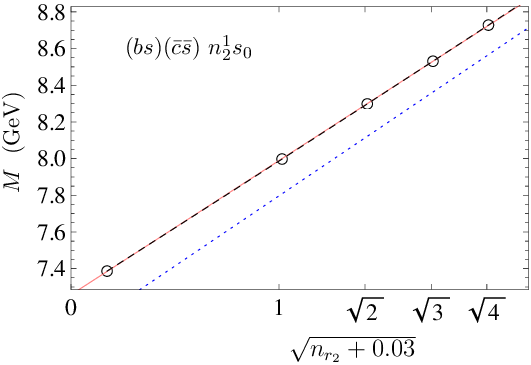}}
\subfigure[]{\label{subfigure:cfa}\includegraphics[scale=0.45]{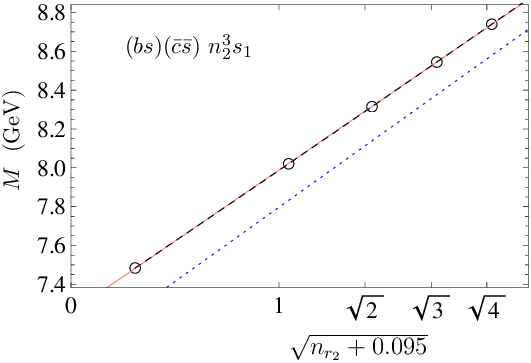}}
\subfigure[]{\label{subfigure:cfa}\includegraphics[scale=0.45]{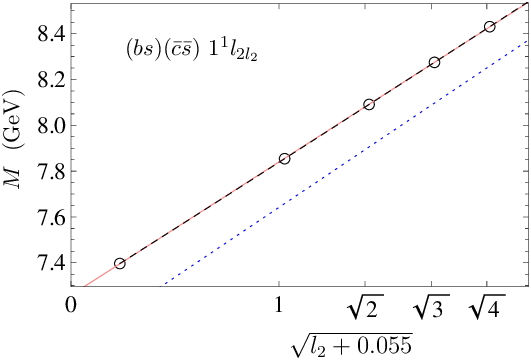}}
\subfigure[]{\label{subfigure:cfa}\includegraphics[scale=0.45]{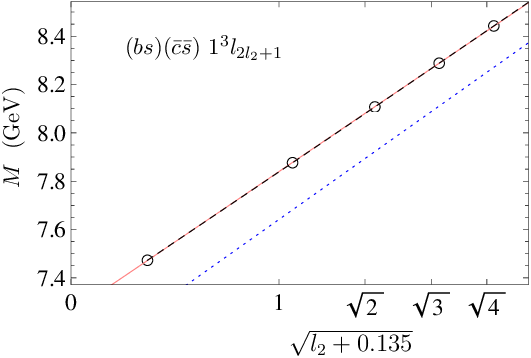}}
\caption{Orbital and radial $\rho_2$-{\trs} for tetraquarks $(bu)(\bar{c}\bar{u})$, $(bu)(\bar{c}\bar{s})$, $(bs)(\bar{c}\bar{u})$, and $(bs)(\bar{c}\bar{s})$. $n_{r_2}$ and ${l_2}$ are the radial and orbital quantum numbers for the $\rho_2$-mode, respectively. Circles represent the predicted data listed in Table \ref{tab:massrhob}. The black dashed lines correspond to complete forms of the $\rho_2$-{\trs}, obtained from Eqs. (\ref{t2qx}) and (\ref{pa2qQx}) or (\ref{summf}) and (\ref{pa2qQx}). The pink lines correspond to the fitted formulas, obtained by linearly fitting the calculated data in Table \ref{tab:massrhob}; these formulas are listed in Table \ref{tab:formulas}. The blue dotted lines are for the main parts of the full forms, which are also listed in Table \ref{tab:formulas}. $n_2^1s_0$ and $n^3_2s_1$ denote radial $\rho_2$-{\trs} for spin-0 and spin-1 antidiquarks, respectively; $1^1l_{2l_2}$ and $1^3l_{2l_{2}+1}$ denote orbital $\rho_2$-{\trs} for spin-0 and spin-1 antidiquarks, respectively.}\label{fig:rhob}
\end{figure*}

For the {\bct} $(bq)(\bar{c}\bar{q}')$, the masses of the radially and orbitally $\rho_2$-excited states are greater than those of the corresponding $\rho_1$-excited states (see Tables \ref{tab:massrho} and \ref{tab:massrhob}). Conversely, for the {\bct} $(cq')(\bar{b}\bar{q})$, the masses of the $\rho_1$-excited states exceed those of the $\rho_2$-excited states. This is because the masse increase from excitation of the diquark $(cq)$ is greater than that from excitation of the diquark $(bq)$.
Accordingly, the following inequalities hold for {\bcts}:
\begin{align}\label{masseq}
M\left(\rho_2, (bq)(\bar{c}\bar{q}')\right)>
M\left(\rho_1, (bq)(\bar{c}\bar{q}')\right),\nonumber\\
M\left(\rho_1, (cq)(\bar{b}\bar{q}')\right)>
M\left(\rho_2, (cq)(\bar{b}\bar{q}')\right),
\end{align}
where $M\left(\rho_1, (bq)(\bar{c}\bar{q}')\right)$ denotes the masses of the radially or orbitally $\rho_1$-mode excited state of $(bq)(\bar{c}\bar{q}')$; the other notations in Eq. (\ref{masseq}) are defined analogously.
The masses of the $\rho_1$-excited ($\rho_2$-excited) states of $(bq)(\bar{c}\bar{q}')$ are equal to those of the $\rho_2$-excited ($\rho_1$-excited) states of $(cq')(\bar{b}\bar{q})$; that is,
\begin{align}
M\left(\rho_1, (bq)(\bar{c}\bar{q}')\right)&=M\left(\rho_2, (cq')(\bar{b}\bar{q})\right),\nonumber\\
M\left(\rho_2, (bq)(\bar{c}\bar{q}')\right)&=M\left(\rho_1, (cq')(\bar{b}\bar{q})\right).
\end{align}

For the light mesons, {\rts} in the $(M^2,x)$ plane are linear and simple, whereas for {\bcts}, they are nonlinear and tedious. For example, the fitted radial $\rho_1$-{\rt} for $(bu)(\bar{c}\bar{u})$ is given by $M=6.91247+ 0.737925 \sqrt{0.125+ n_{r_1}}$ [Eq. (\ref{fra})]. The corresponding expression in terms of $M^2$ becomes
\begin{align}\label{msqr}
M^2=&47.7822+ 10.2018\sqrt{0.125+ n_{r_1}}\nonumber\\
&+0.544533(0.125+ n_{r_1}).
\end{align}
The squared complete form of $\rho$-{\trs} will be even more complicated, for example, see Eq. (\ref{fulla}).
Furthermore, the behavior of the Regge trajectory is less transparent in terms of $M^2$ than in terms of $M$. For the $\rho_1$-{\rt}, $M{\sim}n_{r_1}^{1/2}$, while $M^2{\sim}n_{r_1}^{\nu}$ with $1/2<\nu<1$, where $\nu$ varies for different {\trs}. When Eq. (\ref{msqr}) is approximated as $M^2=m'_R+\beta'(x+c_0)^{1/2}$ due to the dominant role of the $\sqrt{0.125+ n_{r_1}}$ term, the resulting estimates  become less accurate than those obtained using $M=m_R+\beta(x+c_0)^{1/2}$.

Inspired by Ref. \cite{Burns:2010qq}, we employ the $(M,\,(x+c_0)^{\nu})$ plane, rather than the $(M^2,\,x)$ plane, to plot the $\rho_1$- and $\rho_2$-{\trs} for the {\bcts} $(bq)(\bar{c}\bar{q}')$ (see Figs. \ref{fig:rhoa} and \ref{fig:rhob}).
The full forms of the $\rho_1$- and $\rho_2$-{\trs}, directly calculated from Eqs. (\ref{t2qx}) and (\ref{pa2qQx}) or from Eqs. (\ref{summf}) and (\ref{pa2qQx}), are indicated by the black dashed lines in Figs. \ref{fig:rhoa} and \ref{fig:rhob}. For illustration, a representative complete expression is given in Eq. (\ref{fulla}). These full expressions are generally lengthy and cumbersome, and their complexity varies with different {\rts}.
By performing a linear fit to the calculated data in Tables \ref{tab:massrho} and \ref{tab:massrhob}, we obtain the fitted formulas listed in Table \ref{tab:formulas}. In Figs. \ref{fig:rhoa} and \ref{fig:rhob}, the fitted formulas are represented by the pink lines.
We can see that these fitted formulas (pink lines) nearly overlap with the complete forms of the $\rho_1$- and $\rho_2$-{\trs} (black dashed lines). This demonstrates that the complex full forms of the $\rho$-{\rts} can be well approximated by the simple fitted formulas. Accordingly, the complete forms of the $\rho_1$- and $\rho_2$-{\trs} both exhibit the behavior $M{\sim}x_{\rho_1}^{1/2},\;x_{\rho_2}^{1/2}$, where $x_{\rho_1}=n_{r_1},\;l_1$ and $x_{\rho_2}=n_{r_2},\;l_2$.
The main parts of the full forms of the $\rho_1$- and $\rho_2$-{\trs} (listed in Table \ref{tab:formulas} and indicated by the blue dotted lines in Figs. \ref{fig:rhoa} and \ref{fig:rhob}) show significant deviations from the full forms (black dashed lines).

In summary, the Chew-Frautschi plots in Figs. \ref{fig:rhoa} and \ref{fig:rhob} clearly display the trajectory behavior (see the footnote on the first page). The tedious complete forms of the $\rho$-{\trs}--in which the {\rt} behavior is not obvious--can be well approximated by the simple fitted forms, in which the {\rt} behavior is apparent.
The preceding discussions of $\rho$-trajectories for the {\bcts} {\bctf} are identical to those for {\bcts} $(cq')(\bar{b}\bar{q})$.

\subsection{$\lambda$-{\trs}}\label{subsec:rts}

\begin{table*}[!phtb]
\caption{Masses of the $\lambda$-mode excited states of bottom-charm tetraquarks (in ${\gev}$). The notation in Eq. (\ref{tetnot}) is rewritten as $|n_1^{2s_1+1}l_{1j_1},n_2^{2s_2+1}l_{2j_2},N^{2s_3+1}L_J\rangle$. Eqs. (\ref{t2qx}), (\ref{fitcfxl}) and (\ref{fitcfxnr}) are used.}  \label{tab:masslambda}
\centering
\begin{tabular*}{1.0\textwidth}{@{\extracolsep{\fill}}ccccc@{}}
\hline\hline
  $|n_1^{2s_1+1}l_{1j_1},n_2^{2s_2+1}l_{2j_2},N^{2s_3+1}L_J\rangle$        & $(bu)(\bar{c}\bar{u})$  &  $(bu)(\bar{c}\bar{s})$  &  $(bs)(\bar{c}\bar{u})$ &  $(bs)(\bar{c}\bar{s})$\\
\hline
 $|1^1s_0, 1^1s_0, 1^1S_0\rangle$  &7.17    &7.28    &7.28    &7.39 \\
 $|1^1s_0, 1^1s_0, 2^1S_0\rangle$  &7.68    &7.78    &7.79    &7.89  \\
 $|1^1s_0, 1^1s_0, 3^1S_0\rangle$  &8.04    &8.14    &8.15    &8.24  \\
 $|1^1s_0, 1^1s_0, 4^1S_0\rangle$  &8.35    &8.44    &8.46    &8.55  \\
 $|1^1s_0, 1^1s_0, 5^1S_0\rangle$  &8.63    &8.72    &8.74    &8.82  \\
 $|1^1s_0, 1^1s_0, 1^1S_0\rangle$  &7.18    &7.28    &7.29    &7.39 \\
 $|1^1s_0, 1^1s_0, 1^1P_1\rangle$  &7.53    &7.63    &7.64    &7.74 \\
 $|1^1s_0, 1^1s_0, 1^1D_2\rangle$  &7.80    &7.89    &7.91    &8.00 \\
 $|1^1s_0, 1^1s_0, 1^1F_3\rangle$  &8.02    &8.12    &8.13    &8.23  \\
 $|1^1s_0, 1^1s_0, 1^1G_4\rangle$  &8.23    &8.32    &8.34    &8.43 \\
 $|1^1s_0, 1^1s_0, 1^1H_5\rangle$  &8.42    &8.51    &8.53    &8.62 \\
\hline\hline
\end{tabular*}
\end{table*}

\begin{figure*}[!phtb]
\centering
\subfigure[]{\label{subfigure:cfa}\includegraphics[scale=0.45]{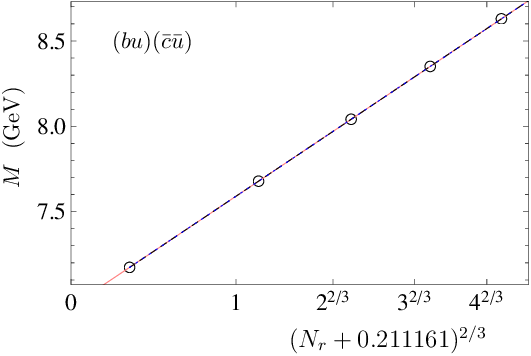}}
\subfigure[]{\label{subfigure:cfa}\includegraphics[scale=0.45]{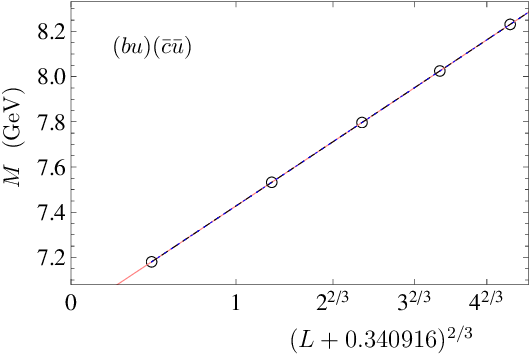}}
\subfigure[]{\label{subfigure:cfa}\includegraphics[scale=0.45]{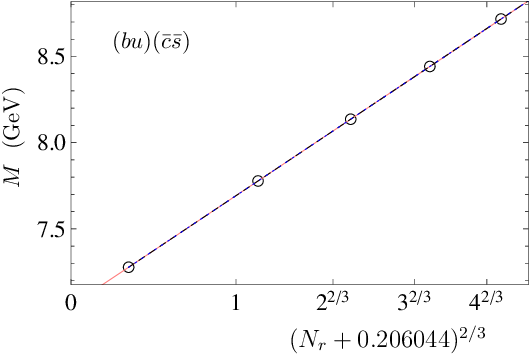}}
\subfigure[]{\label{subfigure:cfa}\includegraphics[scale=0.45]{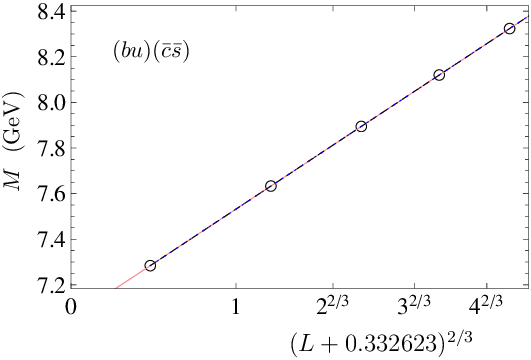}}
\subfigure[]{\label{subfigure:cfa}\includegraphics[scale=0.45]{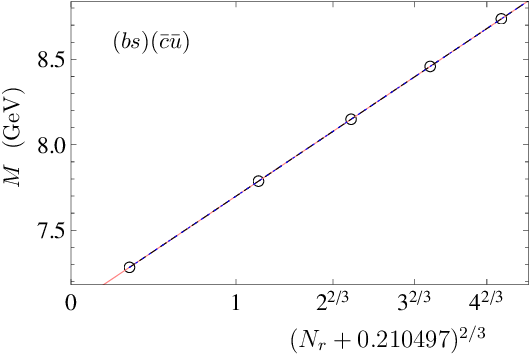}}
\subfigure[]{\label{subfigure:cfa}\includegraphics[scale=0.45]{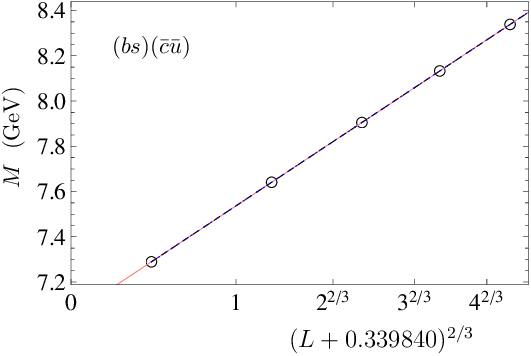}}
\subfigure[]{\label{subfigure:cfa}\includegraphics[scale=0.45]{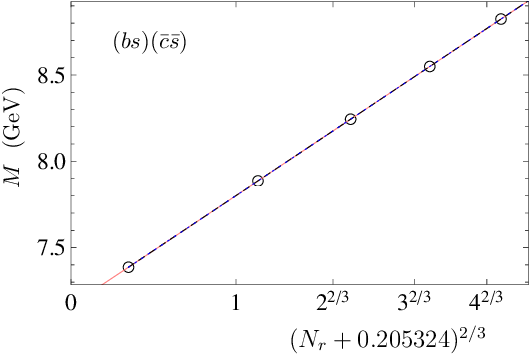}}
\subfigure[]{\label{subfigure:cfa}\includegraphics[scale=0.45]{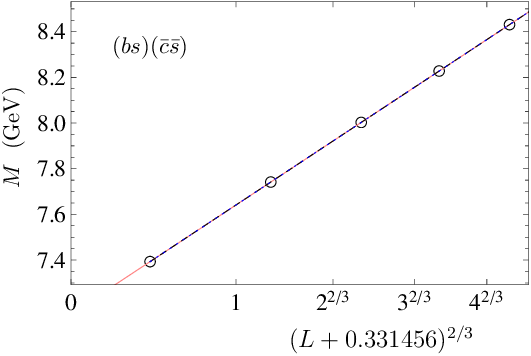}}
\caption{Orbital and radial $\lambda$-{\trs} for tetraquarks $(bu)(\bar{c}\bar{u})$, $(bu)(\bar{c}\bar{s})$, $(bs)(\bar{c}\bar{u})$, and $(bs)(\bar{c}\bar{s})$. $N_{r}$ and ${L}$ are the radial and orbital quantum numbers for the $\lambda$-mode, respectively. Circles represent the predicted data listed in Table \ref{tab:masslambda}. The black dashed lines correspond to the $\lambda$-{\trs} for the complete forms, obtained from Eqs. (\ref{t2qx}) and (\ref{pa2qQx}) or (\ref{summf}) and (\ref{pa2qQx}). The pink lines correspond to the fitted formulas, obtained by linearly fitting the calculated data in Table \ref{tab:masslambda}; these formulas are listed in Table \ref{tab:formulas}. The blue dotted lines are for the main parts of the full forms, which are also listed in Table \ref{tab:formulas}.}\label{fig:lambda}
\end{figure*}

\begin{table*}[!phtb]
\caption{Comparison of theoretical predictions for masses of {\bcts} (in GeV). }  \label{tab:masscomp}
\centering
\begin{tabular*}{1.0\textwidth}{@{\extracolsep{\fill}}cccccccc@{}}
\hline\hline
 $J^{PC}$ &  $|n_1^{2s_1+1}l_{1j_1},n_2^{2s_2+1}l_{2j_2},N^{2s_3+1}L_J\rangle$
          & Tetraquark  &   Our  & EFGL \cite{Ebert:2007rn} &  WLLZ \cite{Wu:2018xdi} & ZH \cite{Zhang:2009vs} & AAS \cite{Agaev:2016dsg}\\
\hline
 $0^{+}$ &  $|1^1s_0, 1^1s_0, 1^1S_0\rangle$ &
                $(bu)(\bar{c}\bar{u})$ &7.17  &7.177  &7.152 &7.12  &7.11  \\
         &   &  $(bu)(\bar{c}\bar{s})$ &7.28  &7.294  &7.248 &  &  \\
         &   &  $(bs)(\bar{c}\bar{u})$ &7.28  &7.282  &7.263 &  &  \\
         &   &  $(bs)(\bar{c}\bar{s})$ &7.39  &7.398  &7.362 &      &7.16  \\
 $1^{+}$ &$|1^3s_1, 1^1s_0, 1^3S_1\rangle$  &
               $(bu)(\bar{c}\bar{u})$  &7.20  &7.198  &7.189 &7.21  & \\
         &   & $(bu)(\bar{c}\bar{s})$  &7.31  &7.317  &7.288 &  & \\
         &   & $(bs)(\bar{c}\bar{u})$  &7.32  &7.302  &7.299 &  & \\
         &   & $(bs)(\bar{c}\bar{s})$  &7.42  &7.418  &7.400 &  & \\
         &$|1^1s_0, 1^3s_1, 1^3S_1\rangle$  &
               $(bu)(\bar{c}\bar{u})$  &7.28  &7.242  &7.283 &7.28  & \\
         &   & $(bu)(\bar{c}\bar{s})$  &7.37  &7.362  &7.374 &  & \\
         &   & $(bs)(\bar{c}\bar{u})$  &7.39  &7.346  &7.390 &  & \\
         &   & $(bs)(\bar{c}\bar{s})$  &7.48  &7.465  &7.465 &  & \\
\hline\hline
\end{tabular*}
\end{table*}

When calculating the $\lambda$-mode excitations, we use scalar diquarks and scalar antidiquarks; other modes are kept in their ground states, and the parameters correspond to the radial ground states of those modes. The radially and orbitally excited $\lambda$-mode states are calculated using Eqs. (\ref{t2qx}), (\ref{fitcfxl}) and (\ref{fitcfxnr}) and the parameters in Table \ref{tab:parmv}. The calculated results are listed in Table \ref{tab:masslambda}. Additional results can be obtained analogously.

Fig. \ref{fig:lambda} shows the radial and orbital $\lambda$-{\trs} for the {\bcts} $(bq)(\bar{c}\bar{q}')$. Circles represent the predicted data (listed in Table \ref{tab:masslambda}), calculated using Eqs. (\ref{t2qx}) and (\ref{pa2qQx}) or Eqs. (\ref{summf}) and (\ref{pa2qQx}). The black dashed lines represent the complete forms of the $\lambda$-{\trs}, obtained from Eqs. (\ref{t2qx}) and (\ref{pa2qQx}) or from Eqs. (\ref{summf}) and (\ref{pa2qQx}). The pink lines represent the fitted formulas, obtained by a linear fit to the calculated data in Table \ref{tab:masslambda} and listed in Table \ref{tab:formulas}. The blue dotted lines correspond to the main parts of the full forms, which are also given in Table \ref{tab:formulas}.
Since the diquark and the antidiquark are treated as individual constituents without considering their internal substructures, the $\lambda$-trajectories are the simplest among the three series of {\rts} (see Table \ref{tab:formulas} and Fig. \ref{fig:lambda}). In Fig. \ref{fig:lambda}, the black dashed lines, pink lines, and blue dotted lines for the $\lambda$-trajectories overlap; that is, the compete forms, fitted formulas, and main parts are identical for these trajectories.
The compete forms of the $\lambda$-trajectories for the {\bcts} $(bq)(\bar{c}\bar{q}')$ behave as $M{\sim}x_{\lambda}^{2/3}$, where $x_{\lambda}=N_r,\,L$.

The preceding discussions of $\lambda$-trajectories for the {\bcts} {\bctf} are identical to those for $(cq')(\bar{b}\bar{q})$.

\subsection{Discussions}
The diquark {\rts} are neither the $\rho_1$-{\trs} for the {\bcts} {\bctf}, nor the main part of the $\rho_1$-{\trs}; see Eqs. (\ref{t2qx}), (\ref{pa2qQx}), (\ref{summf}), (\ref{rtmaina}), and (\ref{rtmainr}). However, the diquark {\rts} play dominant roles in the $\rho_1$-{\trs} and in the main part of the $\rho_1$-{\trs}. A similar statement holds for the antidiquark {\rts} and the $\rho_2$-{\trs}.

One merit of the Regge trajectory approach is its simple analytical form. However, the complete forms of the $\rho_1$- and $\rho_2$-trajectories for {\bcts} are quite lengthy and cumbersome (see, for example, Eq. (\ref{fulla})). Fitting the calculated data yields simplified relations, listed in Table \ref{tab:formulas}. Because the used data are calculated using the complete forms of the {\rts}, the lengthy complete forms can be well approximated by the simple fitted relations; they thus exhibit the same behavior (see Figs. \ref{fig:rhoa} and \ref{fig:rhob}).

It is noteworthy that the main parts of the complete forms of the $\rho$-{\trs} and the fitted formulas share the same functional form, $M=m_R+\beta(x+c_0)^{1/2}$. However, the values of the parameters $m_R$ and $\beta$ differ considerably. 
For example, for the main part of the complete form of a $\rho_1$-{\tr}, $m_R=6.68172$ and $\beta= 0.742965$ [Eq. (\ref{fulla})]; these values are readily computed using Eqs. (\ref{t2qx}) and (\ref{pa2qQx}), or using Eqs. (\ref{summf}) and (\ref{pa2qQx}), and clearly depend on the string tension. By contrast, for the fitted $\rho_1$-{\tr}, $m_R= 6.91247$ and $\beta=0.737925$ [Eq. (\ref{fra})]; here, the dependence on the constituents' masses and the string tension is not readily apparent and becomes more complicated. This indicates that, unlike in the heavy-light meson case, the dependence of $m_R$ on the constituents' masses and the dependence of $\beta$ on the string tension are no longer obvious or direct when we construct simply fitted $\rho_1$- and $\rho_2$-{\tr} formulas for {\bcts}.
As discussed above and in Sec. \ref{subsec:rho}, the $\rho_1$ and $\rho_2$ {\rts} for {\bcts} cannot be obtained by merely mimicking the heavy-light meson {\rts}; instead, they should be constructed based on the actual structure and substructure of the {\bcts}.
Otherwise, the $\rho_1$- and $\rho_2$-trajectories must rely solely on fitting existing theoretical or future experimental data. Consequently, the fundamental relationships between the slopes of the obtained trajectories and string tension, and between $m_R$ and the constituents' masses, will become unobvious and complicated. The predictive power of the Regge trajectories would be compromised.

{\rts} take different forms and behave differently in various energy regions \cite{Chen:2022flh,Chen:2021kfw}.
Both the diquark $(bq)$ and antidiquark $(\bar{c}\bar{q}')$ are the heavy-light systems; therefore, the $\rho$-trajectories of the {\hbts} behave as $M{\sim}x_{\rho}^{1/2}$, see Eq. (\ref{t2qx}). For the $\lambda$-mode, the tetraquark $(bq)(\bar{c}\bar{q}')$ is a heavy-heavy system; hence, the $\lambda$-trajectories of the {\hbts} behave as $M{\sim}x_{\lambda}^{2/3}$.
For {\hbts}, both the $\lambda$-trajectories and the $\rho$-trajectories are concave downwards in the $(M^2,\,x)$ planes, provided the confining potential is linear, regardless of whether the light-quark masses are included, owing to the large heavy-quark masses (see Eq. (\ref{summf})).

To our knowledge, this is the first systematic discussion of all three series of {\rts} for {\bcts} {\bctf} and $(cq)(\bar{b}\bar{q}')$.
Prior work has successfully applied the relation in Eq. (\ref{massform}), the potential in Eq. (\ref{potv}), and the parameter values in Table \ref{tab:parmv} to mesons, baryons, diquarks, triquarks, tetraquarks, and pentaquarks \cite{Chen:2023djq,Chen:2022flh,
Feng:2023txx,Xie:2024lfo,Xie:2024dfe,Song:2024bkj,Song:2025cla}. They are now employed to study {\bcts} {\bctf} and $(cq)(\bar{b}\bar{q}')$, yielding results consistent with other theoretical predictions (see Table \ref{tab:masscomp}). This not only demonstrates the universality of the {\rt} relation and parameter values but also illustrates its predictive capability.

\section{Conclusions}\label{sec:conc}
In this work, we apply the newly proposed tetraquark {\rt} relations to the {\bcts} {\bctf} and $(cq)(\bar{b}\bar{q}')$. We investigate three series of Regge trajectories for the {\bcts}: the $\rho_1$-, $\rho_2$-, and $\lambda$-trajectories. The masses of the $\rho_1$-, $\rho_2$-, and $\lambda$-excited states are roughly estimated.

The complete forms of the Regge trajectories for {\bcts} are often lengthy and cumbersome. Except for the $\lambda$-{\trs}, the $\rho_1$- and $\rho_2$-{\trs} for {\bcts} cannot be obtained simply by mimicking meson {\rts}; instead, they should be constructed based on tetraquark's structure and substructure.
Otherwise, the $\rho_1$- and $\rho_2$-trajectories must rely solely on fitting existing theoretical or future experimental data. The fundamental relationships between the slopes of the obtained trajectories and the string tension, and between $m_R$ and the constituents' masses, would become unobvious. The predictive power of the Regge trajectories would be compromised.

We show that the lengthy complete forms of the $\rho_1$- and $\rho_2$-trajectories can be well approximated by the simple fitted formulas. The $\rho_1$- and $\rho_2$-{\trs} both exhibit the behavior $M{\sim}x_{\rho_1}^{1/2},\;x_{\rho_2}^{1/2}$, where $x_{\rho_1}=n_{r_1},\;l_1$ and $x_{\rho_2}=n_{r_2},\;l_2$. Meanwhile, the $\lambda$-{\trs} exhibit a behavior of $M{\sim}x_{\lambda}^{2/3}$, where  $x_{\lambda}=N_{r},L$.
Moreover, for {\hbts}, all three {\rts} are concave downwards in the $(M^2,\,x)$ plane, provided the confining potential is linear, regardless of whether the light-quark masses are included, owing to the large heavy-quark masses.

\section*{Acknowledgments}
We are very grateful to the anonymous referees for the valuable comments and suggestions.

\vspace{0.3cm}

\noindent{\bf Data Availability Statement} This manuscript has no associated data. [Author's comment: All data are included in the manuscript.]

\vspace{0.3cm}
\noindent{\bf Code Availability Statement} This manuscript has no associated code/software. [Author's comment: The manuscript has no associated code/software.]

\vspace{0.3cm}
\noindent{\bf Open Access} This article is licensed under a Creative Commons Attribution 4.0 International License, which permits use, sharing, adaptation, distribution and reproduction in any medium or format, as long as you
give appropriate credit to the original author(s) and the source, provide a link to the Creative Commons licence, and indicate if changes
were made. The images or other third party material in this article are included in the article's Creative Commons licence, unless indicated
otherwise in a credit line to the material. If material is not included in the article's Creative Commons licence and your intended
use is not permitted by statutory regulation or exceeds the permitted use, you will need to obtain permission directly from the copyright
holder. To view a copy of this licence, visit http://creativecommons.org/licenses/by/4.0/.

\noindent{Funded by SCOAP$^{3}$.}

\appendix
\section{List of the {\rt} relations}\label{sec:appa}
The concrete forms of the {\rts} for tetraquark are not as simple as those for mesons. Although Eqs. (\ref{t2qx}) and (\ref{pa2qQx}) or (\ref{summf}) and (\ref{pa2qQx}) are compact, their final forms can be rather tedious due to tetraquark substructures and the mass dependence of the $\lambda$-trajectory slope.

From Eqs. (\ref{t2qx}) and (\ref{pa2qQx}), or from Eqs. (\ref{summf}) and (\ref{pa2qQx}), we can easily obtain the complete forms of the {\rts} for the {\bcts} {\bctf}. The resulting expressions are rather long and tedious. As an example, we list the radial $\rho_1$-{\tr} for the states $|n_1^1s_0, 1^1s_0, 1^1S_0\rangle$ of tetraquark $(bu)(\bar{c}\bar{u})$, which reads
\begin{widetext}
\begin{align}\label{fulla}
M=&6.68172+0.742965 \sqrt{0.125+n_{r_1}}+0.572052 \Big(\big(6.98172+0.742965 \sqrt{0.125+n_{r_1}}\big)/\big(5.06+0.742965 \sqrt{0.125+n_{r_1}}\big)\Big)^{1/3} \nonumber\\ &\Big(0.334-(0.16719 (5.06+0.742965 \sqrt{0.125+n_{r_1}}))/(6.98172+0.742965 \sqrt{0.125+n_{r_1}})\Big)^{2/3}\nonumber\\
 &\Big(1.008+(0.0153738 (5.06+0.742965 \sqrt{0.125+n_{r_1}}))/(6.98172+0.742965 \sqrt{0.125+n_{r_1}})\Big).
\end{align}
The corresponding main part of the full form (\ref{fulla}) is
\bea\label{fullb}
M=6.68172+0.742965 \sqrt{0.125+n_{r_1}}.
\eea
The fitted formula corresponding to Eq. (\ref{fulla}) is
\bea\label{fra}
M=6.91247+ 0.737925 \sqrt{0.125+ n_{r_1}}.
\eea
\end{widetext}
The $\rho_1$-excited masses for the states $|n_1^1s_0, 1^1s_0, 1^1S_0\rangle$ of tetraquark $(bu)(\bar{c}\bar{u})$ can be calculated using Eq. (\ref{fulla}); the corresponding results are listed in Table \ref{tab:massrho}. By linearly fitting the calculated masses in the $(M,(c0+ x)^1/2)$ plane, we obtain the fitted formulas in Eq. (\ref{fra}). $m_R$ and slope of the fitted {\rt} [Eq. (\ref{fra})] are different from those of the main parts of the complete forms of the {\rt} [see Eq. (\ref{fulla}) or Table \ref{tab:formulas}].

The explicit forms of the {\rts} calculated from Eqs. (\ref{t2qx}) and (\ref{pa2qQx}) or (\ref{summf}) and (\ref{pa2qQx}) are often rather tedious. Here, we present only the main parts of the complete forms of the Regge trajectories, as well as the fitted formulas obtained by fitting the calculated results (see Table \ref{tab:formulas}).

\begin{table*}[!phtb]
\caption{Fitted formulas and main parts of the $\rho$- and $\lambda$-{\rts} for {\bcts} {\bctf}. ${\rm Fit}$: formula obtained by fitting the calculated results; ${\rm Main}$: main part of the complete forms of the {\rts} [derived from Eqs. (\ref{t2qx}) and (\ref{pa2qQx}) or (\ref{summf}) and (\ref{pa2qQx})]. $n_1^1s_0$, $n^3_1s_1$: radial {\rts} for spin-0 and spin-1 diquarks; $1^1l_{1l_1}$, $1^3l_{1l_{1}+1}$: orbital {\rts} for spin-0 and spin-1 diquarks; $n_2^1s_0$, $n^3_2s_1$: radial $\rho_2$-{\trs} for spin-0 and spin-1 antidiquarks; $1^1l_{2l_2}$, $1^3l_{2l_{2}+1}$: orbital $\rho_2$-{\trs} for spin-0 and spin-1 antidiquarks.}  \label{tab:formulas}
\centering
\begin{tabular*}{1.0\textwidth}{@{\extracolsep{\fill}}cccc@{}}
\hline\hline
   &     & Fit  &  Main   \\
\hline
$(bu)(\bar{c}\bar{u})$
    & $\lambda$
        & $M=6.9444+ 0.646189(0.211161+ N_r)^{2/3}$
        & $M=6.9444+ 0.646189(0.211161+ N_r)^{2/3}$      \\
    &   & $M=6.9444+0.483522 (0.340916+ L)^{2/3}$
        & $M=6.9444+0.483522 (0.340916+ L)^{2/3}$       \\
    & $\rho_1$, $n_{1}^1s_0$
        & $M=6.91247+ 0.737925 \sqrt{0.125+ n_{r_1}}$
        & $M=6.681720+ 0.742965 \sqrt{0.125+ n_{r_1}}$       \\
    & $\rho_1$, $n_{1}^3s_1$
        & $M=6.91243+ 0.737954\sqrt{0.155+ n_{r_1}}$
        & $M=6.681720+ 0.742965\sqrt{0.155+ n_{r_1}}$       \\
    &$\rho_1$, $1^1l_{1l_1}$
        & $M=6.91264+ 0.574887\sqrt{0.18+ l_1}$
        & $M=6.681720+ 0.579 \sqrt{0.18+ l_1}$       \\
    &$\rho_1$, $1^3l_{1l_1+1}$
        & $M=6.91261+ 0.574908 \sqrt{0.22+ l_1}$
        & $M=6.681720+ 0.579\sqrt{0.22+ l_1}$       \\
    & $\rho_2$, $n_{2}^1s_0$
        & $M=6.98865+ 0.717668\sqrt{0.065+ n_{r_2}}$
        & $M=6.752678+ 0.751988\sqrt{0.065+ n_{r_2}}$       \\
    & $\rho_2$, $n_{2}^3s_1$
        & $M=6.98672+ 0.718924\sqrt{0.17+ n_{r_2}}$
        & $M=6.752678+ 0.751988\sqrt{0.17+ n_{r_2}}$       \\
    &$\rho_2$, $1^1l_{2l_2}$
        & $M=6.9896+ 0.592938\sqrt{0.095+ l_2}$
        & $M=6.752678+ 0.6228\sqrt{0.095+ l_2}$       \\
    &$\rho_2$, $1^3l_{2l_2+1}$
        & $M=6.98842+ 0.593705\sqrt{0.19+ l_2}$
        & $M=6.752678+ 0.6228\sqrt{0.19+ l_2}$       \\
$(bu)(\bar{c}\bar{s})$
    & $\lambda$
        & $M=7.05501+ 0.637753(0.206044+ N_r)^{2/3}$
        & $M=7.05501+ 0.637753(0.206044+ N_r)^{2/3}$      \\
    &   & $M=7.05501+ 0.477310 (0.332623+ L)^{2/3}$
        & $M=7.05501+ 0.477310 (0.332623+ L)^{2/3}$       \\
    & $\rho_1$, $n_{1}^1s_0$
        & $M=7.01651+ 0.737673 \sqrt{0.125+ n_{r_1}}$
        & $M=6.792332+ 0.742965\sqrt{0.125+ n_{r_1}}$       \\
    & $\rho_1$, $n_{1}^3s_1$
        & $M=7.01646+ 0.737703 \sqrt{0.155+ n_{r_1}}$
        & $M=6.792332+ 0.742965\sqrt{0.155+ n_{r_1}}$       \\
    &$\rho_1$, $1^1l_{1l_1}$
        & $M=6.91264+ 0.574887 \sqrt{0.18+ l_1}$
        & $M=6.792332+ 0.579\sqrt{0.18+ l_1}$       \\
    &$\rho_1$, $1^3l_{1l_1+1}$
        & $M=7.01665+ 0.574705 \sqrt{0.22+ l_1}$
        & $M=6.792332+ 0.579\sqrt{0.22+ l_1}$       \\
    & $\rho_2$, $n_{2}^1s_0$
        & $M=7.149+ 0.731285\sqrt{0.03+ n_{r_2}}$
        & $M=6.922678+ 0.764020\sqrt{0.03+ n_{r_2}}$       \\
    & $\rho_2$, $n_{2}^3s_1$
        & $M=7.1475+ 0.732262\sqrt{0.095+n_{r_2}}$
        & $M=6.922678+ 0.764020\sqrt{0.095+n_{r_2}}$       \\
    &$\rho_2$, $1^1l_{2l_2}$
        & $M=7.14975+ 0.581482\sqrt{0.055+ l_2}$
        & $M=6.922678+ 0.609000\sqrt{0.055+ l_2}$       \\
    &$\rho_2$, $1^3l_{2l_2+1}$
        & $M=7.14872+ 0.582151\sqrt{0.135+ l_2}$
        & $M=6.922678+ 0.609000\sqrt{0.135+ l_2}$       \\
$(bs)(\bar{c}\bar{u})$
    & $\lambda$
        & $M=7.05442+0.645069 (0.210497+N_r)^{2/3}$
        & $M=7.05442+0.645069 (0.210497+N_r)^{2/3}$     \\
    &   & $M=7.05442+0.482697 (0.339840+L)^{2/3}$
        & $M=7.05442+0.482697 (0.339840+L)^{2/3}$       \\
    & $\rho_1$, $n_{1}^1s_0$
        & $M=7.08116+ 0.711933 \sqrt{0.08+ n_{r_1}}$
        & $M=6.851720+ 0.716645\sqrt{0.08+ n_{r_1}}$       \\
    & $\rho_1$, $n_{1}^3s_1$
        & $M=7.08111+ 0.711964 \sqrt{0.11+ n_{r_1}}$
        & $M=6.851720+ 0.716645\sqrt{0.11+ n_{r_1}}$       \\
    &$\rho_1$, $1^1l_{1l_1}$
        & $M=7.0813+ 0.547606\sqrt{0.115+ l_1}$
        & $M=6.851720+ 0.5514\sqrt{0.115+ l_1}$       \\
    &$\rho_1$, $1^3l_{1l_1+1}$
        & $M=7.08126+ 0.54763\sqrt{0.16+l_1}$
        & $M=6.851720+ 0.5514\sqrt{0.16+l_1}$       \\
    & $\rho_2$, $n_{2}^1s_0$
        & $M=7.09786+ 0.717385\sqrt{0.065+ n_{r_2}}$
        & $M=6.862698+ 0.751988\sqrt{0.065+ n_{r_2}}$       \\
    & $\rho_2$, $n_{2}^3s_1$
        & $M=7.09593+ 0.718642 \sqrt{0.17+n_{r_2}}$
        & $M=6.862698+ 0.751988\sqrt{0.17+n_{r_2}}$       \\
    &$\rho_2$, $1^1l_{2l_2}$
        & $M=7.09881+ 0.592703\sqrt{0.095+ l_2}$
        & $M=6.862698+ 0.6228\sqrt{0.095+ l_2}$       \\
    &$\rho_2$, $1^3l_{2l_2+1}$
        & $M=7.09764+ 0.59347\sqrt{0.19+l_2}$
        & $M=6.862698+ 0.6228\sqrt{0.19+l_2}$       \\
$(bs)(\bar{c}\bar{s})$
    & $\lambda$
        & $M=7.16503+0.636602 (0.205324+N_r)^{2/3}$
        & $M=7.16503+0.636602 (0.205324+N_r)^{2/3}$      \\
    &   & $M=7.16503+0.476463 (0.331456+L)^{2/3}$
        & $M=7.16503+0.476463 (0.331456+L)^{2/3}$       \\
    & $\rho_1$, $n_{1}^1s_0$
        & $M=7.18513+ 0.711696 \sqrt{0.08+ n_{r_1}}$
        & $M=6.962332+ 0.716645\sqrt{0.08+ n_{r_1}}$       \\
    & $\rho_1$, $n_{1}^3s_1$
        & $M=7.18508+ 0.711728\sqrt{0.11+ n_{r_1}}$
        & $M=6.962332+ 0.716645\sqrt{0.11+ n_{r_1}}$       \\
    &$\rho_1$, $1^1l_{1l_1}$
        & $M=7.18527+ 0.547416 \sqrt{0.115+ l_1}$
        & $M=6.962332+ 0.5514\sqrt{0.115+ l_1}$       \\
    &$\rho_1$, $1^3l_{1l_1+1}$
        & $M=7.18524+ 0.547441 \sqrt{0.16+ l_1}$
        & $M=6.962332+ 0.5514\sqrt{0.16+ l_1}$       \\
    & $\rho_2$, $n_{2}^1s_0$
        & $M=7.25815+ 0.730998\sqrt{0.03+ n_{r_2}}$
        & $M=7.032698+ 0.764020\sqrt{0.03+ n_{r_2}}$       \\
    & $\rho_2$, $n_{2}^3s_1$
        & $M=7.25665+ 0.731976\sqrt{0.095+ n_{r_2}}$
        & $M=7.032698+ 0.764020\sqrt{0.095+ n_{r_2}}$       \\
    &$\rho_2$, $1^1l_{2l_2}$
        & $M=7.2589+ 0.581253\sqrt{0.055+l_2}$
        & $M=7.032698+ 0.609000\sqrt{0.055+l_2}$       \\
    &$\rho_2$, $1^3l_{2l_2+1}$
        & $M=7.25787+ 0.581923 \sqrt{0.135+l_2}$
        & $M=7.032698+ 0.60900\sqrt{0.135+l_2}$       \\
\hline\hline
\end{tabular*}
\end{table*}

\end{document}